\documentclass[twocolumn,pra,amssymb,eqsecnum]{revtex4}
\usepackage{pslatex}
\usepackage[T1]{fontenc}
\usepackage[latin1]{inputenc}
\usepackage{graphicx}

\makeatletter

\providecommand{\tabularnewline}{\\}

\usepackage{graphics}
\usepackage{amsmath}
\usepackage{epsfig}
\usepackage{bm}

\makeatother
\begin{document}

\title{Gauge Poisson representations for birth/death master equations}

\author{P. D. Drummond}

\email{drummond@physics.uq.edu.au}

\homepage{www.physics.uq.edu.au/BEC}

\affiliation{Universit\" at Erlangen-N\" urnberg, Lehrstuhl f\" ur Optik, Staudstrasse 7/B2 D-91058 Erlangen, Germany.}

\thanks{Permanent Address: Centre for Quantum-Atom Optics, University of Queensland, Brisbane, QLD 4072 Australia}

\date{\today{}}

\begin{abstract}
Poisson representation techniques provide a powerful method for mapping master equations for birth/death processes --- found
in many fields of physics, chemistry and biology --- into more tractable stochastic differential equations. However, the
usual expansion is not exact in the presence of boundary terms, which commonly occur when the differential equations are
nonlinear. In this paper, a gauge Poisson technique is introduced that eliminates boundary terms, to give an exact representation
as a weighted rate equation with stochastic terms. These methods provide novel techniques for calculating and understanding
the effects of number correlations in systems that have a master equation description. As examples, correlations induced
by strong mutations in genetics, and the astrophysical problem of molecule formation on microscopic grain surfaces are analyzed.
Exact analytic results are obtained that can be compared with numerical simulations, demonstrating that stochastic gauge
techniques can give exact results where standard Poisson expansions are not able to.
\end{abstract}
\maketitle

\section{Introduction\label{INTRO} }

The calculation and prediction of the behavior of complex systems is one of the most pressing issues in theoretical physics\cite{Synergetics}
and in many related fields. A common difficulty when dealing with statistical problems is that the state-space of possible
outcomes is enormous. This is particularly so for quantum systems --- but very similar issues can arise in many types of
master equation, with applications ranging from kinetic theory to genetics. One of the earliest approaches to this problem
was the method of Langevin equations in Brownian motion, which led to the theory of equations with random terms, or stochastic
equations. An important subsequent development in the field of discrete master equations was the van Kampen system-size expansion\cite{Kampen},
which leads to an approximate Fokker-Planck equation equivalent to a stochastic equation --- whose deterministic part has
the usual rate-equation behavior. Following this, a more systematic technique was introduced, called the Poisson\cite{Poisson}
expansion. This gives an exact Fokker-Planck equation in cases with linear rate equations, and does not require a system-size
expansion.

The general advantage of Poisson methods is that they employ a `natural' basis in which the distribution is expanded in the
most entropically likely distribution for linear couplings. This allows for very efficient treatment of the underlying Poissonian
statistics. The disadvantage is that Poisson methods involve a complex extension to the usual real space of number densities.
This can result in large errors --- both random and systematic --- when nonlinear interactions generate unstable trajectories
in the complex space. These are solutions to the deterministic drift equations which can reach infinity in a finite time.
Just as in related quantum phase-space methods\cite{DG-PosP}, such trajectories result in power-law distribution tails which
can cause large sampling errors. These can also give rise to systematic boundary term errors\cite{Smith-Gard}, since the
derivation of the Fokker-Planck equation requires that the resulting distribution can be integrated by parts with vanishing
boundary terms. 

In this paper, a new method called the gauge Poisson representation is introduced. This removes any unstable trajectories
or moving singularities, which are conjectured\cite{GGD-Validity} to be the cause of boundary term errors --- thus allowing
an exact mapping of many important master equations into stochastic differential equations, even when the Poisson expansion
cannot be used. This is an example of a stochastic gauge\cite{GaugeP}, in which stochastic equations are modified by introducing
an equivalence class of gauges that stabilize all complex trajectories. I find that correctly chosen stochastic gauges appear
to eliminate the boundary term problem, and simultaneously give rise to greatly reduced sampling errors in practical numerical
solutions. Numerical simulations are presented for specific cases to verify that gauges with no moving singularities lead
to exact results.

To illustrate the technique, I show how the gauge method can be used to calculate means and correlation predictions from
master equations. The first example describes genetic mutations in a simple model from evolutionary biology\cite{eco}. The
second treats the astrophysical problem of interstellar molecular hydrogen production on grain surfaces\cite{Biham}. The
examples show relatively simple behavior that leads to sub-Poissonian results in cases where there is an analytic theory
available to compare with numerical simulations. This demonstrates that correct results can be obtained in the stochastic
gauge simulations, with a range of stabilizing gauges --- even when incorrect results are obtained with the standard Poisson
expansion, due to boundary terms. At the same time, these master equations have a great deal of intrinsic scientific interest.

In many cases, there are more degrees of freedom with correlations and fluctuations that are closer to Poissonian. A type
of problem where stochastic gauge methods would be useful is the treatment of complex systems where there are large numbers
of modes, like a lattice or spatially extended continuum model. Correlations in these types of system are also able to be
treated, in principle, by these methods. Other examples are cases where the statistics that are dominated by some rate-limiting
step involving only small numbers. Details of these applications will be given elsewhere.

\section{Birth-death Master equations}

In master equations, the fundamental object is a probability distribution $P(\mathbf{N},t)=\left[\mathbf{P}(t)\right]_{\mathbf{N}}$
for observing probabilities of discrete outcomes, labeled with integers $\mathbf{N}=(N_{1},\ldots,N_{d})$. These numbers
are typically the number of particles or atoms (in physics), molecules (in chemistry) or organisms (in biology)\cite{Synergetics}.
The numbers may refer to a large, well-mixed volume, or to cells within a larger volume in the case of spatially extended
systems. A common and very significant problem is the Markovian time-evolution of the distribution, defined by an $\mathbf{N}\times\mathbf{N}$
matrix $\mathbf{M}$ so that:

\begin{equation}
\frac{\partial}{\partial t}\mathbf{P}(t)=\mathbf{M}\cdot\mathbf{P}(t)\,\,\,.\label{eq:MarkovME}\end{equation}

There are severe complexity issues that arise in trying to solve this numerically as $d$, the number of modes or dimensions,
increases (unless the problem is exactly soluble or factorisable, which is rarely the case in practice). The difficulty of
solving this equation directly is that even when the maximum number is bounded by $N_{max}$, the total number of discrete
states $\mathbf{N}$ involved is $N_{max}^{d}$, which grows exponentially large with the number of distinct modes $d$.
Thus, direct methods are not suitable for solving many problems of this type. 

The most general Markovian master equation considered here describes a number of coupled reactions, labeled with an index
$a$. Each reaction has the following generic structure:\begin{eqnarray}
\sum_{j}\nu_{j}^{a}X_{j} & \,^{\underrightarrow{k^{a}}}\, & \sum_{j}\mu_{j}^{a}X_{j}\,\,\,.\label{eq:kinetic}\end{eqnarray}

The reactions are restricted to be at most binary, so that $j\le2$, $\mu_{j}^{a},\nu_{j}^{a}=0,1,2$, and $\mu^{a}=\sum_{j}\mu_{j}^{a}\leq2,\,\,\nu^{a}=\sum_{j}\nu_{j}^{a}\leq2$.
The probability of a transition per unit time is proportional to the rate $k^{a}$ and the number of initial `particles',
giving a reaction rate of:\begin{equation}
R^{a}(\mathbf{N})=k^{a}\prod_{j}\frac{N_{j}!}{(N_{j}-\nu_{j}^{a})!}\,\,.\end{equation}

The traditional rate equation for the process (\ref{eq:kinetic}) , which ignores fluctuations, assumes deterministic changes
for $N_{j}\gg1$ according to:

\begin{equation}
\frac{\partial}{\partial t}N_{j}=\sum_{a}(\mu_{j}^{a}-\nu_{j}^{a})R^{a}(\mathbf{N})\,\,.\end{equation}
By contrast, equation (\ref{eq:MarkovME}) describes both mean values and fluctuations. In this case the master equation
for the probability of the outcome $\mathbf{N}$ --- including fluctuations --- has the form:

\begin{equation}
\frac{\partial}{\partial t}P(\mathbf{N})=\sum_{a}\left[R^{a}(\mathbf{N}^{a})P(\mathbf{N}^{a})-R^{a}(\mathbf{N})P(\mathbf{N})\right]\end{equation}
where $N_{j}^{a}=N_{j}+\nu_{j}^{a}-\mu_{j}^{a}$ is the particle number prior to reaction ($a$) that leads to a current
number $N_{j}$. 

The propagation matrix $\mathbf{M}$ can now be constructed from a class of matrix ladder operators which either increase
($\mathbf{L}_{j}^{+}$) the number of particles in a particular mode $j$, or decrease ($\mathbf{L}_{j}^{-}$) the number
of particles and multiply the probability by a factor of $(N_{j}+1)$ . In terms of the probability vector $\mathbf{P}$,
this means that: \begin{eqnarray}
\left[\mathbf{L}_{j}^{+}\mathbf{P}\right]_{\mathbf{N}} & = & P(N_{1},\ldots N_{j}-1,\ldots)\nonumber \\
\left[\mathbf{L}_{j}^{-}\mathbf{P}\right]_{\mathbf{N}} & = & (N_{j}+1)P(N_{1},\ldots N_{j}+1,\ldots)\,\,.\end{eqnarray}

Combining products of these together gives the result that:\begin{equation}
\left[\left(\mathbf{L}_{j}^{+}\right)^{\mu}\left(\mathbf{L}_{j}^{-}\right)^{\nu}\mathbf{P}\right]_{\mathbf{N}}=\frac{(N_{j}+\nu-\mu)!}{(N_{j}-\mu)!}P(N_{1},\ldots N_{j}+\nu-\mu,\ldots)\end{equation}

This is precisely the matrix operation required to construct the master-equation matrix. Hence, after using the identities
for the raising and lowering operators, one finds that the master equation matrix has a factorized structure given by:

\begin{equation}
\mathbf{M}=\sum_{a}k^{a}\left[\prod_{j}\left(\mathbf{L}_{j}^{+}\right)^{\mu_{j}^{a}}-\prod_{j}\left(\mathbf{L}_{j}^{+}\right)^{\nu_{j}^{a}}\right]\left[\prod_{j}\left(\mathbf{L}_{j}^{-}\right)^{\nu_{j}^{a}}\right]\,\,.\label{eq:factor}\end{equation}

\subsection{Poisson Representation}

In this section, the results of the Poisson representation\cite{Poisson} will be summarized. This important development
employs an expansion of the distribution vector $\mathbf{P}$ using `prototype' solutions, namely the complex Poisson distribution
$\mathbf{p}_{0}(\bm\alpha)$, \emph{without requiring a system-size expansion}:\begin{equation}
\left[\mathbf{p}_{0}(\bm\alpha)\right]_{\mathbf{N}}=\prod_{\mathbf{j}=1}^{d}e^{-\alpha_{j}}\left(\alpha_{j}\right)^{N_{j}}/N_{j}!\,\,\,.\end{equation}
The positive Poisson representation\cite{Poisson} expands the distribution vector $\mathbf{P}$ with a quasi-probability,
$f(\bm\alpha)$, defined over a \emph{complex} $d$-dimensional phase-space of variables $\bm\alpha$. \begin{equation}
\mathbf{P}=\int f(\bm\alpha)\mathbf{p}_{0}(\bm\alpha)d^{2d}\bm\alpha\,\,\,.\label{PositiveP}\end{equation}
Here the discrete variable $\mathbf{N}$, which is a vector of integers, is transformed into a continuous variable $\bm\alpha$---
which is a vector of complex numbers. In the above form it is conventional to choose f$(\bm\alpha)$ as positive, so that
it behaves much like a conventional probability. It is also possible to make other choices. For example, the complex Poisson
representation employs a complex contour integral form which is useful for finding exact solutions in special cases. In this
expansion,\begin{equation}
\mathbf{P}=\oint f(\bm\alpha)\mathbf{p}_{0}(\bm\alpha)d^{d}\bm\alpha\,\,\,.\label{ComplexP}\end{equation}

With the aid of differential identities given in the next subsection, either expansion can be used to change the master equation
given above into a differential form. Introducing a generalised measure $d\mu(\bm\alpha)$ to indicate either a volume or
contour integral, and a differential operator $\mathcal{L}'_{P}$ that is determined by the propagation matrix $\mathbf{M}$,
\begin{eqnarray}
\frac{\partial}{\partial t}\mathbf{P}(t) & = & \int f(\bm\alpha)\mathbf{M}\cdot\mathbf{p}_{0}(\bm\alpha)d\mu(\bm\alpha)\,\,\,\nonumber \\
 & = & \int f(\bm\alpha)\left[\mathcal{L}'_{P}\mathbf{p}_{0}(\bm\alpha)\right]d\mu(\bm\alpha)\,\,.\label{anti}\end{eqnarray}

Next, provided that the relevant boundary terms vanish, partial integration results in a modified form with a differential
operator $\mathcal{L}{}_{P}$ acting on the distribution rather than the basis itself, where $\mathcal{L}{}_{P}$ conventionally
is written with all derivative operators on the left:\begin{equation}
\frac{\partial}{\partial t}\mathbf{P}(t)=\int\left[\mathcal{L}{}_{P}f\mathbf{(\bm\alpha)}\right]\mathbf{p}_{0}(\bm\alpha)d\mu(\bm\alpha)\,\,.\label{normal}\end{equation}

This can then be used to deduce that a sufficient condition for the distribution function $f(\bm\alpha)$ is that it should
satisfy a partial-differential equation of Fokker-Planck form:\begin{equation}
\frac{\partial}{\partial t}f\mathbf{(\bm\alpha)}=\mathcal{L}{}_{P}f\mathbf{(\bm\alpha)}\,\,.\label{FPE}\end{equation}
This form is valid in either the positive or the complex Poisson representation. Since a contour integral can be chosen to
be closed, or to have a direction in phase-space which gives rise to an exponentially damped behaviour, it is generally possible
to choose a contour that gives rise to vanishing boundary terms for a complex Poisson representation. The issue is more difficult
in the case of the positive Poisson representation, since the extended (complex) phase-space may include directions where
the Fokker-Planck solutions have power-law tails which are not sufficiently bounded. In such cases, the presence of boundary
terms mean the technique is no longer exact.

In the positive Poisson case, provided the partial differential equation is of second order and has positive-definite diffusion,
an equivalent stochastic differential equation is obtained. The crucial point is that this final equation has just linear
rather than exponential growth in the problem size, as the number of modes $d$ increases.

Observables are calculated using the result that the $m-$th factorial moment is now given by a probabilistic average in
the positive Poisson representation:

\begin{eqnarray}
\langle N_{j}(N_{j}-1)\ldots(N_{j}-m)\rangle & = & \int\alpha_{j}^{m}f(\bm\alpha)d\mu(\bm\alpha)\nonumber \\
 & = & \langle\alpha_{j}^{m}\rangle_{P}\,\,,\label{Stoch-av}\end{eqnarray}
with a similar contour integral result for the complex Poisson representation.

The Poisson expansion is best thought of as providing a systematic procedure that can replace approximations valid for large
enough numbers $N_{j}$, including rate-equations and system-size expansions\cite{Synergetics}. Such large-number approximations
are inapplicable to many important problems where the actual numbers may be small in at least one of the steps. Examples
of this are common in problems involving nano-structures --- like the grains involved in astrophysical molecule production\cite{Biham}.
Other potential applications include genetic population dynamics\cite{eco}, where population numbers in small regions are
also critically important to reproduction, and spatially dependent master equations for diffusion or kinetic processes. Direct
Monte-Carlo simulations can be used in these problems\cite{Gillespie}, but these can be inefficient and time-consuming for
large numbers of modes, since they do not make any use of the fact that most of the populations involved may be nearly Poissonian.

As explained in the Introduction, the Poisson method for birth-death master equations has similar properties to the positive-P\cite{DG-PosP}
representation in quantum mechanics. It is exact if the resulting differential equation is linear, but there are problems
when there are nonlinear terms in the equations. If the Poisson distribution has power-law tails at large radius, then the
resulting transformation develops systematic boundary term errors. These typically develop when unstable trajectories occur\cite{Smith-Gard}
in the drift terms of the corresponding stochastic equations, which can reach the boundary at infinity in a finite time.This
is an intrinsic problem in the derivation of the Fokker-Planck form, Eq (\ref{FPE}), from the earlier integro-differential
equation, Eq (\ref{anti}).

\subsection{Fokker-Planck equation}

For the particular master equations considered here, the Fokker-Planck equation is easily constructed from the matrix factorization
given in Eq (\ref{eq:factor}). The ladder operators obey identities as follows, when acting on a Poisson distribution:\begin{eqnarray}
\mathbf{L}_{j}^{-}\mathbf{p}_{0}(\bm\alpha) & = & \alpha_{j}\mathbf{p}_{0}(\bm\alpha)\nonumber \\
\mathbf{L}_{j}^{+}\mathbf{p}_{0}(\bm\alpha) & = & \left(1+\partial_{j}\right)\mathbf{p}_{0}(\bm\alpha)\,\,.\label{ident}\end{eqnarray}

Since $\mathbf{p}_{0}(\bm\alpha)$ is analytic in $\bm\alpha$, $\bm\partial$ symbolizes either $\left[\partial_{j}^{x}\equiv\partial/\partial x_{j}\right]$
or $-i\left[\partial_{j}^{y}\equiv\partial/\partial y_{j}\right]$ for each of the $j=1,\ldots\,,d$ complex variables $\alpha_{j}=x_{j}+iy_{j}$
. Using the identities of Eq (\ref{ident}), together with Eq (\ref{eq:factor}), one obtains:\begin{equation}
\mathcal{L}'_{P}=\sum_{a}R^{a}(\bm\alpha)\left[\prod_{j}\left(1+\partial_{j}\right)^{\mu_{j}^{a}}-\prod_{j}\left(1+\partial_{j}\right)^{\nu_{j}^{a}}\right]\,\,,\end{equation}
where the Poisson reaction rate $R^{a}(\bm\alpha)$ corresponds to the deterministic reaction rate when $\bm\alpha=\mathbf{N},$
and is given by:\begin{equation}
R^{a}(\bm\alpha)=k^{a}\prod_{j}\alpha_{j}^{\nu_{j}^{a}}\,\,.\end{equation}

Partial integration is used next, in order to obtain a differential operator $\mathcal{L}_{P}$ acting on the distribution
$f$ rather than the expansion kernel $\mathbf{p}_{0}(\bm\alpha)$. This is most conveniently carried out using generalised
spherical coordinates, so that there is one boundary at large radius $r$ , where boundary terms should vanish. More than
one type of generalised radius is possible, and it is useful to define \begin{equation}
r=\sqrt[p]{\sum_{i=1}^{d}\epsilon_{i}|\alpha_{i}|^{p}}\,\,,\label{radius}\end{equation}
where $\epsilon_{i}$ is a multiplicity factor, and $p\ge1$ defines the power law used to obtain the radial norm. Conventional
hyperspherical coordinates are obtained if $p=2$, $\epsilon_{i}=1$. 

Since the kernel can grow as fast as $e^{r}r^{N}$ (for $\Re(\alpha)<0$) , a sufficient condition to have vanishing boundary
terms is that the distribution should vanish faster than $e^{-\lambda r}$, where $\lambda>1$. Since the oscillatory nature
of the kernel for $\Im(\alpha)\neq0$ can cause cancellation of boundary terms, this may not always be necessary. From Eq
(\ref{Stoch-av}), a sufficient condition to obtain a well-defined observable moment of order $m$, is that the distribution
should vanish at $r\rightarrow\infty$ as $r^{-(2d+m)}$ or faster. If the distribution vanishes faster than \textbf{all}
finite power laws, the stochastic equations have a well-defined set of moment equations which are identical to the moment
equations of the original master equation. Provided these have a unique solution for a given initial condition, the less
stringent condition that all moments exist is presumably sufficient to ensure that boundary terms vanish at $r\rightarrow\infty$.

After partial integration (provided boundary terms vanish) the following Fokker-Planck equation is found:

\begin{eqnarray}
\frac{\partial f(\bm\alpha)}{\partial t} & = & \mathcal{L}_{P}f\label{eq:PoissonFPE}\\
 & = & \sum_{a}\left[\prod_{j}\left(1-\partial_{j}\right)^{\mu_{j}^{a}}-\prod_{j}\left(1-\partial_{j}\right)^{\nu_{j}^{a}}\right]R^{a}f(\bm\alpha)\,\,,\nonumber \end{eqnarray}

In cases of interest involving at most binary kinetics, only first and second order derivatives occur, giving a differential
operator in the form:

\begin{equation}
\mathcal{L}_{P}=A_{j}^{+}\partial_{j}+\frac{1}{2}D_{ij}^{+}\partial_{i}\partial_{j}\,\,.\end{equation}
 Here the repeated Latin indices $i,j$ are summed over $i=1,\ldots,d$, so that $A_{j}^{+}$ is a $d$-component complex
vector called the drift vector, while $D_{ij}^{+}$ is a $d\times d$ square complex symmetric matrix called the diffusion
matrix. 

The basic drift and diffusion matrices\cite{Poisson} are given on inspection of the Fokker-Planck equation (\ref{eq:PoissonFPE}),
on considering all possible values and combinations of $\mu_{j}^{a},\nu_{j}^{a}$:\begin{eqnarray}
A_{j}^{+} & = & \sum_{a}\left(\mu_{j}^{a}-\nu_{j}^{a}\right)R^{a}(\bm\alpha)\nonumber \\
D_{ij}^{+} & = & \sum_{a}\left[\mu_{i}^{a}(\mu_{j}^{a}-\delta_{ij})-\nu_{j}^{a}(\nu_{j}^{a}-\delta_{ij})\right]R^{a}(\bm\alpha)\nonumber \\
\end{eqnarray}

\subsection{Stochastic Equations}

To obtain stochastic equations, it is necessary to take a matrix square root to generate the $d\times d'$ noise matrix $\mathbf{B}$,
where \begin{equation}
\mathbf{D}^{+}=\mathbf{BB}^{T}\,\,.\end{equation}
 The lack of uniqueness of matrix square-roots allows arbitrary functions in phase-space to be introduced, called diffusion
gauges\cite{Plimak,GaugeP}. As an example of this, it is always possible to choose a diffusion gauge corresponding to separate
matrices for each different reaction $a$, so that:\begin{equation}
\mathbf{D}^{+}=\sum_{a}R^{a}\mathbf{B^{a}B}^{aT}\,\,.\label{distinct-noise}\end{equation}
With this choice, the noises are always proportional to $\sqrt{R^{a}}$. Since each reaction has an individual noise matrix
of size $d\times d^{a}$, the total noise dimension $d'$ is given by $d'=\sum_{a}d^{a}$. If required, it is also possible
to increase the noise dimension to $d'=2d+\sum_{a}d^{a}$, by adding $d$ matrix terms $\mathbf{B}^{j}$ which have nonzero
entries only in the $j-th$ row :\begin{equation}
\mathbf{B}^{j}=\frac{g_{j}^{D}(\bm\alpha)}{\sqrt{2}}\left[\begin{array}{cccc}
0 & \ldots & 0 & 0\\
\vdots & \ldots & \vdots & \vdots\\
0 & \ldots & 1 & i\\
\vdots & \ldots & \vdots & \vdots\\
0 & \ldots & 0 & 0\end{array}\right]\,\,.\\
\nonumber \end{equation}
These have the property that $\mathbf{B}^{j}\mathbf{B}^{jT}=0$, so therefore they do not alter the diffusion matrix. In
addition to these gauges that change the noise dimension, it is also possible to use orthogonal transformations on $\mathbf{B}$
which keep the dimension invariant, but alter the noise correlations.

The Fokker-Planck differential operator acting on the distribution $f$ is then transformed into a stochastic differential
equation by taking advantage of the equivalent analytic forms in the differential operators, as described in more detail
in the next section. The result is an Ito stochastic differential equation:\begin{eqnarray}
\frac{d\alpha_{j}}{dt} & = & A_{j}^{+}(\bm\alpha)+B_{jk}\zeta_{k}(t)\nonumber \\
 & = & A_{j}^{+}(\bm\alpha)+\sum_{a}\sqrt{R^{a}}B_{jk}^{a}\zeta_{k}(t)+g_{j}^{D}(\bm\alpha)\xi_{j}(t)\,\,,\label{PoissonIto}\end{eqnarray}
where the functions $\zeta_{k}(t)$ are delta-correlated Gaussian real noise terms, with:\begin{equation}
\langle\zeta_{i}(t)\zeta_{j}(t')\rangle=\delta_{ij}\delta(t-t')\,\,.\end{equation}
The stochastic functions $\xi_{k}(t)$ are delta-correlated Gaussian complex noise terms, which give rise to a gauge symmetry,
in that they have no effect on the resulting moments:\begin{eqnarray}
\langle\xi_{i}(t)\xi_{j}^{\ast}(t')\rangle & = & \delta_{ij}\delta(t-t')\nonumber \\
\langle\xi_{i}(t)\xi_{j}(t')\rangle & = & 0\,\,.\end{eqnarray}

The difficulty with the positive Poisson method\cite{Poisson} outlined above is that even normally stable drift equations
can become unstable due to movable singularities in this extended complex phase-space, which become accessible when the noise
term develops a complex part. The standard term `movable singularity'\cite{Solitons} describes any solution which can reach
infinity in a finite time, depending on the initial conditions. The singularity therefore `moves' with the initial conditions,
rather than occurring at a fixed time. 

These singular trajectories themselves are rare, and may form a set of measure zero on the extended phase-space. However,
just one singularity has been shown in an earlier study of typical examples to lead to Fokker-Planck equations with power-law
tails, that do not vanish sufficiently quickly at the phase-space boundaries\cite{GGD-Validity}. This leads to systematic
boundary term errors in the results, as well as greatly increased numerical integration and sampling errors.

\subsection{Unary reactions}

As an illustration, I will consider some elementary types of reaction, to demonstrate the derivation given here. These results
are readily generalized to the stochastic gauge case treated later.

First, consider one-species reactions --- this can be a simple isomerization or cell diffusion with a rate $\gamma,$ of
the form:

\begin{equation}
X_{1}\,^{\underrightarrow{\gamma}}\, X_{2}\,\,.\end{equation}
In this case, the master-equation reaction matrix has the usual property that the probability of an event is proportional
to the rate and the initial number of particles $N_{x}$. The master equation is therefore:\begin{eqnarray}
\frac{\partial}{\partial t}P(\mathbf{N}) & = & \gamma N_{1}^{+}P(N_{1}^{+},N_{2}^{-})\nonumber \\
 &  & -\gamma N_{1}P(\mathbf{N})\,\,,\end{eqnarray}
where $N_{j}^{\pm}=N_{j}\pm1$. This can also be represented using matrices as:\begin{equation}
\frac{\partial}{\partial t}\mathbf{P}=\gamma\left[\mathbf{L}_{2}^{+}-\mathbf{L}_{1}^{+}\right]\mathbf{L}_{1}^{-}\mathbf{P}\,\,.\end{equation}
In this case the corresponding differential operator is:\begin{equation}
\mathcal{L}_{P}'=\gamma\alpha_{1}\left[\partial_{2}-\partial_{1}\right]\,\,.\end{equation}

On transforming into a Fokker-Planck equation, one obtains: \begin{equation}
\frac{\partial}{\partial t}f\mathbf{(\bm\alpha)}=\gamma\left[\partial_{1}-\partial_{2}\right]\alpha_{1}f\mathbf{(\bm\alpha)}\,\,.\label{FPE0}\end{equation}
Hence, the deterministic differential equation for the characteristics, which are noise-free in this case, are:\begin{eqnarray}
\frac{d\alpha_{1}}{dt} & = & -\gamma\alpha_{1}\nonumber \\
\frac{d\alpha_{2}}{dt} & = & \gamma\alpha_{1}\,\,.\end{eqnarray}

The important advantage of the Poisson method is that these equations have an identical form to simple rate-equations, yet
they are exact, and include all the relevant statistics. Since the Green's function is a delta-function, an initially bounded
distribution remains bounded, and there are no boundary terms.

\subsection{Dimerization}

To illustrate the procedure in nonlinear cases, consider a dimerization process:\begin{equation}
X_{1}+X_{1}\,^{\underrightarrow{k}}\, X_{2}\,\,.\end{equation}
The master equation can be represented using the elementary matrix operators as:\begin{equation}
\frac{\partial}{\partial t}\mathbf{P}=k\left[\mathbf{L}_{2}^{+}-\mathbf{L}_{1}^{+}\mathbf{L}_{1}^{+}\right]\mathbf{L}_{1}^{-}\mathbf{L}_{1}^{-}\mathbf{P}\,\,.\end{equation}
The corresponding differential operator is:\begin{equation}
\mathcal{L}_{P}'=k\alpha_{1}^{2}\left[\partial_{2}-2\partial_{1}-\partial_{1}^{2}\right]\,\,.\end{equation}

As long as partial integration is permissible (which is questionable here) the Fokker-Planck equation would be:\begin{equation}
\frac{\partial}{\partial t}f\mathbf{(\bm\alpha)}=k\left[2\partial_{1}-\partial_{2}-\partial_{1}^{2}\right]\alpha_{1}^{2}f\mathbf{(\bm\alpha)}\,\,.\label{FPE1}\end{equation}
Hence, the corresponding stochastic differential equations are:\begin{eqnarray}
\frac{d\alpha_{1}}{dt} & = & -2k\alpha_{1}^{2}+i\alpha_{1}\sqrt{2k}\zeta(t)\nonumber \\
\frac{d\alpha_{2}}{dt} & = & k\alpha_{1}^{2}\,\,,\end{eqnarray}
where $\langle\zeta(t)\zeta(t')\rangle=\delta(t-t')$ . 

Only the first equation needs to considered in detail, as it is autonomous. In this case, on defining dimensionless variables
$\tau=2kt$, $n=\alpha_{1},$ the ungauged Poisson equation reduces to the form\begin{equation}
\frac{dn}{d\tau}=-n^{2}+in\eta(\tau)\,\,,\label{dimerizeP}\end{equation}
with $\langle\eta(\tau)\eta(\tau')\rangle=\delta(\tau-\tau')$. 

In recent studies of these equations, clear evidence was found of substantial numerical errors\cite{Deloub}. To understand
this, note that there is a movable singularity in this drift equation of the form $n(\tau)=1/(\tau-\tau_{0})$. In stochastic
calculations, it is found that random trajectories are generated for negative initial conditions (due to the noise term),
and these can be arbitrarily close to the singularity. 

To show the analytic consequences of the singularity\cite{Smith-Gard}, consider the inverse variable $z=1/n$, which has
the linear Ito stochastic equation:

\begin{equation}
\frac{dz}{d\tau}=1-z-iz\eta(\tau)\,\,.\label{dimerizePinv}\end{equation}
This equation has a uniform noiseless flow at $z=0$, with no absorbing submanifold. Hence it has a continuous distribution
without a zero in the inverse variable distribution $f_{inv}(z)$. On transforming back to the original variables, the Jacobean
of the transformation generates a power law tail, with $f(n)\propto1/|n|^{4}=1/r^{4}$. From the earlier analysis of boundary
terms, this power-law tail is insufficient to ensure the existence of any distribution moments. One must therefore expect
systematic errors due to boundary terms which do not vanish on partial integration.

\subsection{Generic binary reactions}

As a final illustration, consider a generic binary interaction, in which two species are transformed at a rate $k$ into
two new species:

\begin{equation}
X_{1}+X_{2}\,^{\underrightarrow{k}}\, X_{3}+X_{4}\,\,.\end{equation}
The master equation is:\begin{eqnarray}
\frac{\partial}{\partial t}P(\mathbf{N}) & = & k(N_{1}^{+})(N_{2}^{+})P(N_{1}^{+},N_{2}^{+},N_{3}^{-},N_{4}^{-})\nonumber \\
 &  & -kN_{1}N_{2}P(\mathbf{N})\,\,.\end{eqnarray}
This can also be represented using matrices as:\begin{equation}
\frac{\partial}{\partial t}\mathbf{P}=k\left[\mathbf{L}_{3}^{+}\mathbf{L}_{4}^{+}-\mathbf{L}_{1}^{+}\mathbf{L}_{2}^{+}\right]\mathbf{L}_{1}^{-}\mathbf{L}_{2}^{-}\mathbf{P}\,\,.\end{equation}
Hence, in this case:\begin{eqnarray}
\mathcal{L}'_{P} & = & k\alpha_{1}\alpha_{2}\left[\left(1+\partial_{3}\right)\left(1+\partial_{4}\right)-\left(1+\partial_{1}\right)\left(1+\partial_{2}\right)\right]\nonumber \\
 & = & k\alpha_{1}\alpha_{2}\left[\partial_{3}+\partial_{4}-\partial_{1}-\partial_{2}+\partial_{3}\partial_{4}-\partial_{1}\partial_{2}\right]\,\,.\label{LA}\end{eqnarray}

In the present example, this procedure --- which is only valid if boundary terms vanish during the partial integration ---
would result in a stochastic equation for the complex Poisson mean variables $\alpha_{j}$. On introducing the complex `reaction
rate', $R=k\alpha_{1}\alpha_{2}$, the resulting Ito equations are: \begin{eqnarray}
\frac{d\alpha_{1}}{dt} & = & -R+i\sqrt{R/2}\left(\zeta_{1}+i\zeta_{2}\right)\nonumber \\
\frac{d\alpha_{2}}{dt} & = & -R+i\sqrt{R/2}\left(\zeta_{1}-i\zeta_{2}\right)\nonumber \\
\frac{d\alpha_{3}}{dt} & = & R+\sqrt{R/2}\left(\zeta_{3}+i\zeta_{4}\right)\nonumber \\
\frac{d\alpha_{4}}{dt} & = & R+\sqrt{R/2}\left(\zeta_{3}-i\zeta_{4}\right)\,\,.\label{PSDE}\end{eqnarray}

This shows very clearly a useful property of the Poisson method. The modified particle statistics caused by nonlinear reactions
are immediately apparent from the noise terms, since any fluctuations represent a departure from Poisson statistics. 

As in the previous example, singular trajectories can occur at negative values of $\alpha$, which can be reached via stochastic
motion in the complex plane. Singularities like this often exist in complex nonlinear equations of polynomial form, since
these systems are generically non-integrable or even chaotic --- and the Painleve conjecture\cite{Painleve} states that
movable singularities are to be expected in analytically continued non-integrable sets of equations. Thus, in this case also
the boundary terms may not vanish, leading to systematic errors. Although such errors are known to be exponentially small
when there is large linear damping, they can cause problems when there is little or no linear damping.

\subsection{Numerical Simulations}

The results of direct numerical simulations can be used to test the accuracy of a stochastic method. The numerical results
included in this paper are simple examples where the detailed results of simulations can be evaluated in exactly soluble
cases.

\subsubsection{Stratonovich calculus}

Stochastic calculus is normally carried out in one of two different forms. The first is the Ito calculus, where all terms
that multiply a stochastic noise are evaluated before carrying out the stochastic step forward in time. This is the simplest
form, and corresponds directly to the coefficients in the type of Fokker-Planck equation used elsewhere in this section.
The second is the Stratonovich form, where all the multiplicative terms are evaluated implicitly at the midpoint of a given
step forward in time. This form corresponds to taking the wide-band limit of a finite band-width stochastic equation, and
follows more standard calculus rules for variable-changes.

For numerical simulations\cite{Comp} it is generally more efficient to use Stratonovich equations\cite{Poisson} --- in
which the Ito drift term is modified in a standard way to allow central difference algorithms to be employed. In a generic
Ito equation like Eq (\ref{PoissonIto}), the Stratonovich method generates a modified drift term $A_{j}^{s}$, where:\begin{equation}
A_{j}^{s}=A_{j}-\frac{1}{2}B_{ik}\partial_{i}B_{jk}\,\,.\end{equation}
The resulting equations can be used directly in stable implicit central-difference algorithms, which are robust and well-suited
to the present nonlinear equations. Here the (possible) ambiguity in the analytic differential notation is immaterial, since
by construction the noise matrix $B_{jk}$ is analytic or meromorphic.

\subsubsection{Error estimation}

Discretization error can be estimated by comparing simulations with different step-sizes, but identical underlying noise
sources. This error was typically of order $10^{-3}$ in the simulations in this paper. The algorithm used was an iterative
implicit central difference method\cite{Comp}, which directly implements the Stratonovich form of the stochastic equation.
All numerical code was generated in C++ including estimators of both the sampling error and discretization error, using an
XML script and an automatic code generator\cite{xmds} obtained from the XMDS project web-site.

As an estimator of sampling error, I use the Gaussian estimator of the standard deviation in the mean, $\sigma_{g}=\sigma/\sqrt{N_{s}}$.
However, more sophisticated estimators must be used when the results are strongly non-Gaussian. To ensure that the results
were a strong test of the stochastic gauge method, a large number of samples ($N_{s}=10^{6}$ ) were used in the numerical
calculations reported here, so as to give low sampling errors $\sigma_{g}$.

\section{Gauge Poisson Representation}

As shown in the previous section, the positive Poisson method may have systematic errors in cases involving nonlinear drift,
due to boundary terms on partial integration caused by unstable trajectories. The gauge Poisson representation introduced
here treats the problem of boundary terms, by utilizing a gauge technique similar to that recently proposed for the positive-P
distribution\cite{GaugeP}. It adds an extra variable to the distribution, which eliminates instabilities by modifying the
dynamical equations. A type of gauge-invariance allows this to be carried out exactly. The gauge equations retain the advantages
of the Poisson method, but have no boundary term errors for suitably chosen gauges. This is essential for correct results.

\subsection{Gauge phase-space expansion}

The technical details are as follows. Define an extended (gauge) phase-space with $\overrightarrow{\alpha}=(\Omega,\,\bm\alpha)$,
and a weighted Poisson distribution as $\mathbf{p}(\overrightarrow{\alpha})=\mathbf{p}_{0}(\bm\alpha)\Omega$. Here $\Omega=\alpha_{0}$
is a complex-valued weighting factor which weights (or multiplies) the usual normalized Poisson basis vector. The gauge expansion
is defined for a real, positive distribution $G(\overrightarrow{\alpha})$, as:\begin{equation}
\mathbf{P}=\int G(\overrightarrow{\alpha})\mathbf{p}(\overrightarrow{\alpha})d^{2}\Omega d^{2d}\bm\alpha\,\,\,.\end{equation}
I will show that this implies that a freedom of choice becomes available in the equivalent stochastic equations. Importantly,
it then is possible to choose an equivalent stochastic equation of motion without instabilities or boundary terms. 

As with the standard Poisson representation, time-evolution is treated here by introducing differential identities, so that:

\begin{eqnarray}
\frac{\partial}{\partial t}\mathbf{P}(t) & = & \int G(\overrightarrow{\alpha})\mathbf{M}\cdot\mathbf{p}(\overrightarrow{\alpha})d^{2}\Omega d^{2d}\bm\alpha\,\,\,\nonumber \\
 & = & \int G(\overrightarrow{\alpha})\left[\mathcal{L}'_{G}\mathbf{p}(\overrightarrow{\alpha})\right]d^{2}\Omega d^{2d}\bm\alpha\,\,\,.\label{anti-g}\end{eqnarray}
Just as previously, the goal of the transformation is to allow partial integration, so that, provided the boundary terms
vanish, this equation has an equivalent form of:

\begin{equation}
\frac{\partial}{\partial t}\mathbf{P}(t)=\int\left[\mathcal{L}{}_{G}G\mathbf{(\overrightarrow{\alpha})}\right]\mathbf{p}(\overrightarrow{\alpha})d^{2}\Omega d^{2d}\bm\alpha\,\,\label{normal-g}\end{equation}

However, the crucial distinction between this method and the usual Poisson method is that the introduction of a weighting
factor in the basis means that there are now additional identities available. This allows the differential operator $\mathcal{L}_{G}$
to be chosen so that the resulting time-evolution of the distribution $G\mathbf{(\overrightarrow{\alpha})}$ remains sufficiently
compact at all times to guarantee that boundary terms vanish.

So far this is similar to the positive Poisson representation\cite{Poisson}. However, the $m-$th factorial moment is now
given by a weighted average, with $\Omega$ as a complex weighting parameter in the averages:

\begin{eqnarray}
\langle N_{j}(N_{j}-1)\ldots(N_{j}-m)\rangle & = & \int\left(\Omega\alpha_{j}^{m}\right)G(\overrightarrow{\alpha})d^{2}\Omega d^{2d}\bm\alpha\nonumber \\
 & = & \langle\Omega\alpha_{j}^{m}\rangle=\langle\langle\alpha_{j}^{m}\rangle\rangle\,\,\,,\label{Stoch-av-g}\end{eqnarray}
where the notation $\langle\langle\ldots.\rangle\rangle$ for a complex Poisson variable means a weighted \emph{stochastic
gauge average}. From this one obtains the expected result that in a pure Poisson distribution with $G(\overrightarrow{\alpha})=\delta(\Omega-1)\prod_{j=1}^{d}\delta(\alpha_{j}-\bar{\alpha}_{j}\,)$,
the mean and variance of modes with $j>0$ are given by:

\begin{eqnarray}
\langle N_{j}\rangle & = & \bar{\alpha}_{j}\nonumber \\
\langle\left(N_{j}-\bar{N}_{j}\right)^{2}\rangle & = & \bar{\alpha}_{j}\,\,\,.\label{eq:variance}\end{eqnarray}

\subsection{Gauge identities}

The extra variable $\Omega$ allows an additional differential identity to be used to introduce a \emph{stochastic gauge}
--- an arbitrary vector function in the extended phase-space with $d+1$ complex dimensions. This can be used to stabilize
the drift equations throughout the extended phase-space, thus allowing integration by parts. There is no free lunch here,
however! The price paid is that there is a new stochastic equation in $\Omega$, leading to a finite variance in the gauge
amplitude $\Omega$. While this can cause practical problems due to sampling errors --- which must be minimized --- it is
important to note that these errors can be estimated and controlled by choice of gauge and by increasing the number of sample
trajectories. By contrast, there is no presently known technique of estimating and controlling boundary term errors in the
standard Poisson expansion. 

The additional identity in the weight variable $\Omega$ has the simple form of:

\begin{equation}
\mathbf{p}(\overrightarrow{\alpha})=\Omega\partial_{\Omega}\mathbf{p}(\overrightarrow{\alpha})\,\,\,.\end{equation}
To derive the stochastic gauge equations, I now introduce $d'$ arbitrary complex \emph{drift gauge} functions $\mathbf{g}=(\, g_{i}(\overrightarrow{\alpha},t)\,)$,
to give a new differential operator $\mathcal{L}'_{G}$ which is equivalent to the usual Poisson operator $\mathcal{L}'_{P}$,
but which includes $\Omega$ derivatives in the extended phase-space: \begin{equation}
\mathcal{L}'_{G}=\mathcal{L}'_{P}+\left[\frac{1}{2}\mathbf{g}\cdot\mathbf{g}\,\Omega\,\partial_{0}+\sum_{j=1}^{d}\sum_{k=1}^{d'}g_{k}B_{jk}\partial_{j}\right]\left[\Omega\partial_{0}-1\right]\,\,\,.\label{eq:Gauge-operator}\end{equation}
To simplify notation, I have used $\partial_{0}$ to symbolize either $\left[\partial_{0}^{x}\equiv\partial/\partial x_{0}\right]$
or $-i\left[\partial_{0}^{y}\equiv\partial/\partial y_{0}\right]$ for the complex weight variable $\Omega=\alpha_{0}=x_{0}+iy_{0}$
. This allows a choice of analytic derivatives, which is later used to obtain a positive definite Fokker-Planck equation.
Since the added term has a factor $\left[\Omega\partial_{0}-1\right]$ which vanishes when operating on the gauge basis $\mathbf{p}(\overrightarrow{\alpha})$,
the gauge functions can be arbitrary, just as in the analogous situation of gauge field symmetries in electrodynamics. 

Summing repeated Latin indices from now on over $i=0,\ldots,d$, Eq (\ref{eq:Gauge-operator}) becomes: \begin{equation}
\mathcal{L}'_{G}=\left[A_{i}\partial_{i}+\frac{1}{2}D_{ij}\partial_{i}\partial_{j}\right]\,\,\,.\end{equation}
 Here, the \emph{total} complex drift vector, including gauge corrections, is $\underline{A}=(0,\, A_{1},\ldots A_{d})$,
where:\begin{equation}
A_{j}=A_{j}^{+}-\sum_{k=1}^{d'}g_{k}B_{jk}\,\,\,\,[j,k>0]\,.\label{gaugedrift}\end{equation}
 This remarkable result shows that as long as there is a non-vanishing noise term, the drift equation can be modified in
an arbitrary way by adding a gauge term. 

The diffusion matrix changes as well. The \emph{total} diffusion matrix $\underline{\underline{D}}$ is a $(d+1)\times(d+1)$
matrix, with a new $(d+1)\times d'$ square root $\underline{\underline{B}}$:

\begin{eqnarray}
\underline{\underline{D}} & = & \left[\begin{array}{cc}
\Omega^{2}\mathbf{gg}^{T}, & \Omega\mathbf{gB}^{T}\\
\mathbf{Bg}^{T}\Omega, & \mathbf{BB}^{T}\end{array}\right]\nonumber \\
 & = & \left[\begin{array}{c}
\Omega\mathbf{g}\\
\mathbf{B}\end{array}\right]\left[\Omega\mathbf{g}^{T},\mathbf{B}^{T}\right]=\underline{\underline{B}}\,\underline{\underline{B}}^{T}\,\,\,.\end{eqnarray}
Thus, the $(d+1)\times d'$ complex stochastic noise matrix $\underline{\underline{B}}$ is as before, except with one added
row:\begin{equation}
\underline{\underline{B}}=\left[\begin{array}{c}
\Omega\mathbf{g}\\
\mathbf{B}\end{array}\right]\,\,\,.\label{gauge_{d}}\end{equation}

The additional row means that whenever a gauge term is added, a corresponding noise term appears in the equation of motion
for the gauge amplitude variable $\Omega$. The details of this are derived next.

\subsection{Stochastic gauge equations}

So far, there is no restriction on which of the choices of analytic derivative is utilized to obtain the identities. This
means that it is possible to use the free choice of equivalent identities to give a differential operator which is entirely
real and has a positive-definite diffusion. This procedure is also followed in the positive-P\cite{DG-PosP} and positive
Poisson\cite{Poisson} representations. Here it is extended\cite{GaugeP} to include the gauge variable $\Omega$ as well
as the other variables. This is achieved by introducing a $2(d+1)$ dimensional real phase space $(x_{0},y_{0},\ldots x_{d},y_{d})$,
with derivatives ${\partial_{\mu}}$, and separating $\underline{\underline{B}}=\underline{\underline{B}}^{x}+i\underline{\underline{B}}^{y}$
into its real and imaginary parts. A similar procedure is followed for $\underline{A}=\underline{A}^{x}+i\underline{A}^{y}$.

The choice for the analytic derivative, where $\partial_{i}\rightarrow\partial_{i}^{x}$ or $\partial_{i}\rightarrow-i\partial_{i}^{y}$,
can now be made definite by choosing it so the resulting drift and diffusion terms are always real. In more detail, this
corresponds to choosing:\begin{eqnarray}
A_{i}\partial_{i} & \rightarrow & A_{i}^{x}\partial_{i}^{x}+A_{i}^{y}\partial_{i}^{y}\,,\\
D_{ij}\partial_{i}\partial_{j} & \rightarrow & B_{ik}^{x}B_{jk}^{x}\partial_{i}^{x}\partial_{j}^{x}+B_{ik}^{y}B_{jk}^{x}\partial_{i}^{y}\partial_{j}^{x}+(x\leftrightarrow y)\,.\nonumber \end{eqnarray}
 At this point it is necessary to introduce a corresponding real drift vector $\mathcal{A}_{\mu}$ and diffusion matrix $\mathcal{D}_{\mu\nu}$
which are defined on the $2(d+1)$ dimensional real phase space. Hence, the gauge differential operator can now be written
explicitly in this equivalent real form, as:\begin{equation}
\mathcal{L}'_{G}=\left[\mathcal{A}_{\mu}\partial_{\mu}+\frac{1}{2}\mathcal{D}_{\mu\nu}\partial_{\mu}\partial_{\nu}\right]\,\,\,,\end{equation}
 where $\underline{\underline{\mathcal{D}}}=\underline{\underline{\mathcal{B}}}\underline{\underline{\mathcal{B}}}^{T}$
is now positive semi-definite. This can be seen by writing $\underline{\underline{\mathcal{B}}}$ as a $2(d+1)\times d'$
real matrix:\begin{equation}
\underline{\underline{\mathcal{B}}}=\left[\begin{array}{c}
\underline{\underline{B}}^{x}\\
\underline{\underline{B}}^{y}\end{array}\right]\,\,\,,\end{equation}
so that the diffusion matrix is the square of a real matrix, with:

\begin{equation}
\underline{\underline{\mathcal{D}}}=\left[\begin{array}{c}
\underline{\underline{B}}^{x}\\
\underline{\underline{B}}^{y}\end{array}\right]\,\times\left[\left(\underline{\underline{B}}^{x}\right)^{T},\left(\underline{\underline{B}}^{y}\right)^{T}\right]\,\,\,.\end{equation}

Hence, choosing the analytic derivatives to give real terms in $\mathcal{L}_{G}$ generates a positive semi-definite diffusion
operator on a real space of $2(d+1)$ dimensions. Provided that one can integrate by parts, the full evolution equation is
then:\begin{eqnarray}
\frac{\partial}{\partial t} & \mathbf{P}(t)= & \int\left[\mathcal{L}{}_{G}G(\overrightarrow{\alpha})\right]\mathbf{p}(\overrightarrow{\alpha})d^{2(d+1)}\overrightarrow{\alpha}\,\,\,.\label{GaugeInt}\end{eqnarray}
 Provided that one can integrate by parts, there is at least one solution for $G$ which satisfies the positive-definite
Fokker-Planck equation: \begin{equation}
\frac{\partial}{\partial t}G(\overrightarrow{\alpha},t)=\left[-\partial_{\mu}\mathcal{A}_{\mu}+\frac{1}{2}\partial_{\mu}\partial_{\nu}\mathcal{D}_{\mu\nu}\right]G(\overrightarrow{\alpha},t)\,\,\,.\label{RealFPE}\end{equation}

It is important to note that the crucial partial integration step is only permissible if the distribution is strongly enough
bounded at infinity ($|\overrightarrow{\alpha}\textrm{|}\rightarrow\infty$) so that all boundary terms vanish. Just as in
the positive-P expansion, this means that the distribution must be bounded in phase-space more strongly than all power laws
in $1/r$ as $r\rightarrow\infty$, in order for the moments to be defined. There is an additional requirement that the distribution
vanishes faster than $1/|\Omega|^{2}$ as $|\Omega|\rightarrow\infty,$ since there is now an additional integration over
$d^{2}\Omega$ to be carried out.

However, the freedom to choose a gauge means that there are now ways to eliminate movable singularities from the drift equations
corresponding to $\mathcal{A}_{\mu}$. I will show in examples given later that this removes boundary terms as well --- as
expected from earlier conjectures about the relation between boundary terms and drift singularities. 

The positive-definiteness of the diffusion matrix $\underline{\underline{\mathcal{D}}}$ implies that the Fokker-Planck equation
is equivalent to a set of $d+1$ Ito stochastic differential equations, with $d'$ \emph{real} Gaussian processes $\zeta_{i}(t)$.
This central result can be written compactly using the complex variable form, as:\begin{eqnarray}
\frac{d\Omega}{dt} & = & \Omega g_{k}\zeta_{k}(t)\,\,,\nonumber \\
\frac{d\alpha_{j}}{dt} & = & A_{j}^{+}(\bm\alpha)+B_{jk}[\zeta_{k}(t)-g_{k}]\,\,\,.\label{eq:GSDE}\end{eqnarray}
 The noises $\zeta_{i}$ have correlations $\langle\zeta_{i}(t)\zeta_{j}(t')\rangle=\delta_{ij}\delta(t-t')$, and are uncorrelated
between time steps. Repeated noise indices are summed over $k=1,d'$.

As with the Poisson representation, if the Stratonovich method is used, a modified drift term $\mathcal{A}_{\mu}^{s}$ is
generated where:\begin{equation}
\mathcal{A}_{\mu}^{s}=\mathcal{A}_{\mu}-\frac{1}{2}\mathcal{B}_{\rho\nu}\partial_{\rho}\mathcal{B}_{\mu\nu}\,\,.\end{equation}
The resulting equations can be used directly in stable implicit central-difference algorithms. Care should be used here in
differentiating the noise matrix $\mathcal{B}_{\mu\nu}$. Since this includes the gauge, and is no longer an analytic function,
the real and imaginary parts need to be treated separately.

\section{Asymptotics and boundary terms}

It is crucial to choose the drift gauge $\mathbf{g}$ so that the resulting distribution is more strongly bounded than any
power-law in the radius, in order to remove boundary terms and ensure that all of the moments are well-defined. Amongst the
gauges that achieve this goal, it is preferable to use one that minimises the sampling error. An empirical rule is that no
deterministic trajectory can be allowed to reach the boundary in a finite time, even on a set of initial conditions with
measure zero, as this is the signature\cite{GGD-Validity} for a distribution with a power-law tail --- which cannot be integrated
by parts exactly, and has large sampling errors. 

However, this is not always a sufficient condition, since it does not take into account the radial dependence of the stochastic
noise. In the generic binary reaction equation (\ref{PSDE}), it is clear that noise term has at most linear radial growth,
since $\sqrt{2|\alpha_{1}\alpha_{2}|}\le|\alpha_{1}|+|\alpha_{2}|\le\epsilon_{1}|\alpha_{1}|+\epsilon_{2}|\alpha_{2}|\le r$,
where $r$ is defined as in Eq (\ref{radius}). More generally, in all cases studied here, the noise has radial components
with no more than linear radial growth in $B_{rj}$. In some cases, the radial noise can vanish. In general, this relatively
slow growth in radial noise means that moments remain well-defined as long as there is no more than linear asymptotic growth
in the radial drift.

\subsection{Minimal gauges}

A second criterion of practical significance, is to use a gauge that generates an attractive subspace on which the drift
gauge vanishes. These gauges are called minimal gauges. If this condition is not satisfied, then the stochastic noise in
the gauge amplitude $\Omega$ creates a relatively large and growing sampling error. This is not just an issue of mathematics,
but also one of computational efficiency. In performing a numerical calculation, there are always some numerical errors.
These are due to the finite nature of computers (and even human calculators). 

Thus, one has to estimate and minimize numerical errors due to round-off error, finite step-size in time, and sampling error
due to the use of a finite sample of trajectories. In general, there is an optimum gauge which minimizes sampling errors,
but even a non-optimal gauge can be used simply by increasing the number of sampled trajectories.

\subsection{Existence of gauges}

In this section, I demonstrate the existence of stabilizing gauges for systems with deterministically stable rate equations.
In later sections, the numerical simulation of a realistic nonlinear master equation --- which generates boundary term errors
\emph{without} a stabilizing gauge --- is shown to give correct results within the sampling error when suitable gauges are
used. The important issue, as always, is that the calculated result must agree with the correct value within a known error-bar.

The gauge must be chosen to stabilise the nonlinear drift equations. Just as in the simpler example of Eq (\ref{LA}), the
drift equations can only have constant, linear and quadratic terms. Hence, the unmodified drift equation for the $i$-th
component can always be written as:\begin{equation}
\frac{\partial\alpha_{i}}{\partial t}=A_{i}^{(0)}+\sum_{j>0}A_{ij}^{(1)}\alpha_{j}+\sum_{j,k>0}A_{ijk}^{(2)}\alpha_{j}\alpha_{k}\,\,\,,\label{eq:quadratic}\end{equation}
where all coefficients are real. 

Any deterministic instability for standard rate equations --- in the subspace of real, positive $\alpha_{j}$ --- is generally
ruled out by number conservation laws \emph{a priori}. In this positive subspace, there is often a conservation law such
that $r=\sum_{i}\epsilon_{i}\alpha_{i}$ is conserved, and hence $A_{r}=0$. Here $\epsilon_{i}$ must be chosen appropriately,
for example as the number of atoms in a given chemical species. The present equations are not always strictly number-conserving
however, since reservoirs are allowed. Nevertheless, I assume them deterministically stable, in the sense that in the positive
subspace, the asymptotic Stratonovich drift must be bounded so that $A_{r}\le ar$ . This does permit exponential growth,
which certainly occurs in many cases.

A movable singularity \emph{can} occur in the analytic continuation of the rate equations, which are the precise equations
found in the usual positive Poisson equations. It is then essential to add a stabilizing gauge that removes any singularities
if the resulting stochastic equations are to be accurate and useful --- although the equations with singularities can still
be used as approximate equations for large particle number. 

The Painleve conjecture\cite{Painleve} states that movable singularities are a generic property of the analytic continuations
of nonlinear sets of equations, so it is to be expected that these will generally occur in quadratic equations of the form
given in Eq (\ref{eq:quadratic}). Each singularity has the signature that for at least one component $j$, it evolves to
$|\alpha_{j}|=\infty$ in a finite time $t_{0}$, and hence must typically have a leading term with an inverse power-law
time-dependence with power $p_{j}$ for $t<t_{0}$:\begin{equation}
\alpha_{j}=\frac{\alpha_{j}^{0}}{(t_{0}-t)^{p_{j}}}\,\,\,.\end{equation}

It is necessary to demonstrate the existence of gauge choices that eliminate these singularities. I now wish to demonstrate
that stabilizing gauges always exist, provided that the deterministic rate equations are stable. This proof gives only minimal
conditions for gauge-stabilization. Other stable gauges also exist, and may be more efficient in terms of sampling error
in any given case.

\subsubsection{Amplitude gauge}

First, it is important to choose an appropriate diffusion gauge --- that is, the choice of matrix square root of the diffusion
matrix must be specified. This is most simply done by choosing to regard each different unidirectional reaction as having
a distinct diffusion term proportional to the reaction rate, as in Eq (\ref{distinct-noise}). Unidirectional reactions are
classified according to the number of initial species and final species. This leads to nine reaction types, as shown in Table
(\ref{cap:Different-types-of}), ignoring factors of order unity for simplicity.

\begin{table}
\begin{center}\begin{tabular}{|c|c|c|c|c|}
\hline 
Initial $\nu$&
Final $\mu$&
Rate $R^{a}$&
Initial Diffusion &
Final Diffusion \tabularnewline
\hline
\hline 
$0$&
$0$&
$k$&
$0$&
$0$\tabularnewline
\hline 
$1$&
$0$&
$k\alpha$&
$0$&
$0$\tabularnewline
\hline 
$2$&
$0$&
$k\alpha_{1}\alpha_{2}$&
$R^{a}$&
$0$\tabularnewline
\hline 
$0$&
$1$&
$k$&
$0$&
$0$\tabularnewline
\hline 
$1$&
$1$&
$k\alpha$&
$0$&
$0$\tabularnewline
\hline 
$2$&
$1$&
$k\alpha_{1}\alpha_{2}$&
$R^{a}$&
$0$\tabularnewline
\hline 
$0$&
$2$&
$k$&
$0$&
$R^{a}$\tabularnewline
\hline 
$1$&
$2$&
$k\alpha$&
$0$&
$R^{a}$\tabularnewline
\hline 
$2$&
$2$&
$k\alpha_{1}\alpha_{2}$&
$R^{a}$&
$R^{a}$\tabularnewline
\hline
\end{tabular}\end{center}

\caption{Different types of uni-directional reactions, classified by initial and final species numbers.\label{cap:Different-types-of}}
\end{table}

Diffusion terms occur only in reaction channels involving two particles. However, in all channels it is always possible to
add a diffusion gauge so that the noise matrix is nonvanishing, from Eq (\ref{PoissonIto}). This diffusion gauge choice
may not always be optimal for sampling purposes, but it is a possible choice, and there is a resulting drift gauge which
is stable.

Defining a phase angle $\phi_{i}$ via: \begin{equation}
\alpha_{i}=|\alpha_{i}|e^{i\phi_{i}}\,\,,\end{equation}
 every drift term has the form \begin{eqnarray}
\frac{d\alpha_{i}}{dt} & = & A_{i}^{+}\nonumber \\
 & = & A_{i}^{(0)}+A_{ij}^{(1)}|\alpha_{j}|e^{i\phi_{j}}+A_{ijk}^{(2)}|\alpha_{j}\alpha_{k}|e^{i(\phi_{j}+\phi_{k})}\,\,.\end{eqnarray}
 It is always possible to subtract the gauge term\begin{eqnarray}
\sum_{j'}B_{ij'}g_{j'}{} & = & A_{ij}^{(1)}\alpha_{j}[1-e^{i(\phi_{i}-\phi_{j})}]\nonumber \\
 &  & +A_{ijk}^{(2)}\alpha_{j}\alpha_{k}[1-e^{i(\phi_{i}-\phi_{j}-\phi_{k})}]\,\,,\end{eqnarray}
which cancels the original rate and replaces it by one at a phase angle $\phi_{i}$ equal to the phase of $\alpha_{i}$.
The deterministic part of such an equation can only modify the amplitude of $\alpha_{i}$, and so is effectively restricted
to a $d$-dimensional real space, just as the usual deterministic rate equations are. But these equations have an asymptotic
 linear radial bound by hypothesis. This gauge is therefore a stabilizing gauge.

An example of this is for $r=k\alpha^{2}$, as in the dimerization equation (\ref{dimerizeP}) which has singular trajectories
in the standard Poisson method. In this case, one can simply choose:

\begin{equation}
gB=A^{+}[1-|\alpha|/\alpha]=-k\alpha[\alpha-|\alpha|]\,\,.\label{amplitudegauge}\end{equation}
The gauged drift equation becomes:\begin{eqnarray}
d\alpha/dt & = & A^{+}-gB\nonumber \\
 & = & -k\alpha|\alpha|\,\,,\end{eqnarray}
which is clearly stable because the drift is directed toward the origin at all times, so $A_{r}\le0$ .

Hence, the gauge corrected equations are stable. When the rate-equations have an attractor, this gauge tends to produce random
circular paths of constant amplitude, instead of localized behavior in the complex phase-space. I will therefore refer to
it as the `amplitude' gauge.

\subsubsection{Phase gauge}

The amplitude gauge can be improved by modifying the gauge term so that it also stabilizes the phase near $\phi_{i}=0$.
This reduces the size of the gauge-induced noise, and hence reduces sampling errors. A suitable choice is to add additional
gauge terms of form: \begin{equation}
\sum_{j'}B_{ij'}g_{j'}{}=i\alpha_{i}a_{i}(\bm\phi,|\bm\alpha|)\,\,.\end{equation}

Here the real function $a(\bm\phi,|\bm\alpha|)$ is defined to generate an attractor at $\bm\phi=0$, where $a(\bm\phi,|\bm\alpha|)=0$,
so the gauge corrections all vanish at zero phase. 

As before, we can consider the dimerization equation (\ref{dimerizeP}), with $rR=-k\alpha^{2}$. In this situation, one
can simply choose $a(\vec{\phi})=ky/\alpha$, where $\alpha=x+iy$, giving an overall gauge contribution of:

\begin{equation}
gB=-k\alpha[\alpha-|\alpha|]+ik\alpha y=-k\alpha[x-|\alpha|]\,\,.\label{phasegauge}\end{equation}
The gauged drift equation then becomes:\begin{equation}
d\alpha/dt=-k\alpha(|\alpha|+iy)\,\,.\end{equation}
The additional term has no effect on global stability, but increases the likelihood of trajectories near $\bm\phi=0$, since
the phase gauge above generates a deterministic equation in the form $d\phi/dt\propto-\phi$. 

As the most likely trajectory is in-phase and has zero gauge correction, this gauge is minimal. Correspondingly, the gauge
noise and resulting sampling errors are reduced, as I will show later in the numerical examples. This gauge will be called
the `phase' gauge, as it stabilizes the phase-angle of $\alpha$ as well as the modulus. Although similar nonlinearities
occur in spatially extended systems\cite{Spatial}, more subtle gauge choices may be better in these cases.

\section{Genetic mutation master equation}

To demonstrate how the positive Poisson method can be usefully employed, consider the important genetic problem of a stochastic
master equation for the evolution of a finite population, where the $N_{j}$ are simply the populations of genotype $j$.
A simple model for linear evolution through asexual reproduction and mutation is of the form\cite{eco}:\begin{eqnarray}
X_{j} & \,^{\underrightarrow{k_{j}}}\, & 0\nonumber \\
X_{i} & \,^{\underrightarrow{k_{ij}}}\, & X_{i}+X_{j}\,\,\,.\label{eq:Gene}\end{eqnarray}
Here $k_{ij}$ is the birth rate, and $k_{j}$ is the death rate. It is sometimes convenient to also define $k_{i}^{b}=\sum_{j}k_{ij}$
as the total birth rate and $Q_{i}$ as the mutation rate, where $k_{ii}=(1-Q_{i})k_{i}^{b}$ . Defining $\mathbf{N}^{\pm}[i]=(N_{1},N_{i}\pm1\ldots,N_{d})$,
the corresponding master equation is:\begin{eqnarray}
\frac{d}{dt}P(\mathbf{N}) & = & -\left(\sum_{i,j}k_{ij}N_{i}+\sum_{i}k_{i}N_{i}\right)P(\mathbf{N})\nonumber \\
 &  & +\sum_{i}k_{i}(N_{i}+1)P(\mathbf{N}^{+}[i])\nonumber \\
 &  & +\sum_{i,j}k_{ij}(N_{j}-1)P(\mathbf{N}^{-}[j])\,\,.\label{eq:Genemaster}\end{eqnarray}

This has well-known problems: the state-space may be very large, preventing a direct matrix solution. On the other hand,
while the corresponding average rate-equations reduce to the widely-studied Eigen\cite{eco} quasi-species model, the rate-equations
are unable to treat population fluctuations in small samples.

An interesting exactly soluble case involves two species, with $Q=1$, so that reproduction always leads to mutation. This
has rates given by:\begin{eqnarray}
k_{1} & =k_{2} & =k\nonumber \\
k_{12} & =k_{21} & =k_{m}\nonumber \\
k_{11} & =k_{22} & =0\,\,.\label{Twospecies}\end{eqnarray}

\subsection{Stochastic equations}

The equivalent Ito stochastic equation is exact for all master equations of the form of Eq (\ref{eq:Genemaster}). It is:\begin{equation}
\frac{d\alpha_{j}}{dt}=-k_{j}\alpha_{j}+\sum_{i}k_{ij}\alpha_{i}+\sum_{k}B_{jk}\zeta_{k}(t)\,\,\,.\end{equation}

Here, the noise matrix $B_{jk}$ is determined from the symmetrized diffusion matrix, $D_{ij}=\left[\alpha_{i}k_{ij}+\alpha_{j}k_{ji}\right]$,
where: \begin{equation}
\sum_{k}B_{ik}B_{jk}=D_{ij}\,\,\,.\end{equation}

This leads immediately an important result: an initially Poissonian distribution is invariant under pure decay processes,
but can develop increased fluctuations with non-Poissonian features due to birth and mutation events, described by the matrix
$D_{ij}$. In general, constructing the square root of a symmetric real matrix $D_{ij}$ is non-unique. The most powerful
technique requires a matrix diagonalization through an orthogonal transformation, and may result in eigenvalues of either
sign. If all the eigenvalues are positive, the resulting fluctuations are super-Poissonian. If some are negative, at least
one sub-Poissonian feature will occur, and a complex stochastic process will result.

A simple example is obtained by considering the symmetric two-species case of Eq (\ref{Twospecies}). In this case, the diffusion
is entirely off-diagonal, and the Poisson representation exactly transforms a complicated master equation into a soluble
stochastic equation. The Ito equations can be simplified further as follows, on introducing population sum and difference
variables $n_{\pm}=(\alpha_{1}\pm\alpha_{2})$ and $k_{\pm}=(k\mp k_{m})$ :\begin{eqnarray}
\frac{dn_{+}}{dt} & = & -k_{+}n_{+}+\sqrt{2k_{m}n_{+}}\zeta_{1}(t)\nonumber \\
\frac{dn_{-}}{dt} & = & -k_{-}n_{-}+i\sqrt{2k_{m}n_{+}}\zeta_{2}(t)\,\,.\label{Mutate2}\end{eqnarray}

In this situation, there is an absorber at $n_{+}=0$, since any stochastic trajectory that reaches this value has a zero
derivative. This is due to the randomness of birth or death events, which mean that it is always possible for the random
walk in this low-dimensional population space to finish at extinction. In addition, any initial differences between the two
populations decays, and is replaced by a strong sub-Poissonian correlation. This is due to the fact that all births must
occur in a way that tends to equalize the two species that are present.

\subsection{Means and correlations}

To calculate analytic solutions, standard Ito calculus can be used to obtain the exact time-evolution of correlations and
expectation values. This is straightforward, since in Ito calculus the noise terms are not correlated with the other stochastic
variables at the same time, so $\langle f(\alpha_{i})\zeta_{j}(t)\rangle=0$. Thus, the means and variances are all soluble
from their respective time-evolution equations --- which is also possible using the master-equation form. 

Equations for general correlations $\overline{\alpha_{l}\alpha_{j}}=\langle\alpha_{l}\alpha_{j}\rangle_{P}$ can be either
calculated from the master equation, or from the stochastic equations. Defining $\Delta k_{ij}=k_{ij}-\delta_{ij}k_{i}$
and using the rules of Ito calculus, one obtains:\begin{eqnarray}
\frac{d\overline{\alpha_{j}}}{dt} & = & \sum_{i}\Delta k_{ij}\overline{\alpha_{i}}\nonumber \\
\frac{d\overline{\alpha_{l}\alpha_{j}}}{dt} & = & \sum_{i}\Delta k_{ij}\overline{\alpha_{l}\alpha_{i}}+\sum_{i}\Delta k_{il}\overline{\alpha_{i}\alpha_{j}}+\overline{D_{lj}}\nonumber \\
\label{Moment}\end{eqnarray}
In the symmetric two-species case, the initial means and correlations are defined as: $\overline{n_{+}}=\langle n_{+}(0)\rangle_{P}$,
$\overline{n_{-}}=\langle n_{-}(0)\rangle_{P}$, $\overline{n_{+}^{2}}=\langle n_{+}^{2}(0)\rangle_{P}$, $\overline{n_{-}^{2}}=\langle n_{-}^{2}(0)\rangle_{P}$
and $\overline{n_{+}n_{-}}=\langle n_{+}(0)n_{-}(0)\rangle_{P}$. Solving the moment equations (\ref{Moment}) gives the
following exact results:\begin{eqnarray}
\langle n_{+}(t)\rangle_{P} & = & \overline{n_{+}}e^{-k_{+}t}\nonumber \\
\langle n_{-}(t)\rangle_{P} & = & \overline{n_{-}}e^{-k_{-}t}\nonumber \\
\langle n_{+}^{2}(t)\rangle_{P} & = & \overline{n_{+}^{2}}e^{-2k_{+}t}+2k_{m}e^{-3k_{+}t/2}\overline{n_{+}}\frac{\sinh(k_{+}t/2)}{k_{+}/2}\nonumber \\
\langle n_{-}^{2}(t)\rangle_{P} & = & \overline{n_{-}^{2}}e^{-2k_{-}t}+\left(\frac{2k_{m}\overline{n_{+}}}{k+3k_{m}}\right)\left(e^{-k_{+}t}-e^{-2k_{-}t}{}\right)\nonumber \\
\langle n_{+}(t)n_{-}(t)\rangle_{P} & = & \overline{n_{+}n_{-}}e^{-2kt}\,\,\,.\label{Genesoln}\end{eqnarray}

Suppose that birth and death rates are equal ($k_{m}=k$), so the mean population $\langle n_{+}(t)\rangle_{P}$ is time-invariant.
Then the asymptotic population differences have a mean and variance of:

\begin{eqnarray}
\lim_{t\rightarrow\infty}\langle N_{-}(t)\rangle & = & 0\nonumber \\
\lim_{t\rightarrow\infty}\langle[N_{-}(t)]^{2}\rangle & = & \lim_{t\rightarrow\infty}\langle n_{-}^{2}(t)+n_{+}(t)\rangle_{P}\nonumber \\
 & = & \frac{\bar{n}_{+}}{2}\,\,.\label{Limit}\end{eqnarray}

This indicates that the population difference has a steady-state variance of half its usual value, owing to the fact that
all births are correlated between the species, thus causing sub-Poissonian statistics --- while all deaths are uncorrelated,
tending to restore the Poissonian distribution. By comparison, the variance in the total population shows linear growth in
this case, as some populations in the total ensemble become extinct, while others can randomly grow to a large population. 

In this case there are no unstable trajectories, and the positive Poisson method can be used directly. However, it should
be noted that this model ignores inter-species competition --- which could lead to nonlinear effects involving boundary terms.

\subsection{Numerical results}

Applying the Stratonovich rules to generate equations for numerical stochastic integration results in:

\begin{eqnarray}
\frac{dn_{+}}{dt} & = & -k_{m}/2-k_{+}n_{+}+\sqrt{2k_{m}n_{+}}\zeta_{1}(t)\nonumber \\
\frac{dn_{-}}{dt} & = & -k_{-}n_{-}+i\sqrt{2k_{m}n_{+}}\zeta_{2}(t)\,\,.\label{MutateStrat}\end{eqnarray}

This illustrates the typical feature of the Stratonovich calculus, which is the generation of terms in the drift equations
due to the noise. Some care is needed in calculations near the absorbing boundary at $n_{+}=0$, which is treated by imposing
an appropriate boundary condition at this point.

\begin{figure}
\includegraphics[%
  width=6cm,
  keepaspectratio]{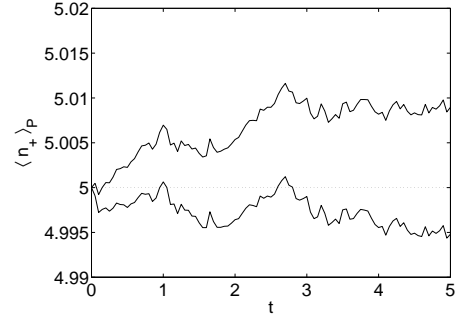}

\caption{Sampled mean populations $\langle n_{+}\rangle_{P}$ for genetic mutation example in the Poisson representation, parameters
as in text, showing upper and lower one standard deviation error bounds. Sampling error ($\sigma_{m}$) is of order $10^{-3}$or
less. These results agree with the analytic theory (dotted line) within the sampling error.\label{cap:Sampled-mean-populations}}
\end{figure}

\begin{figure}
\includegraphics[%
  width=6cm,
  keepaspectratio]{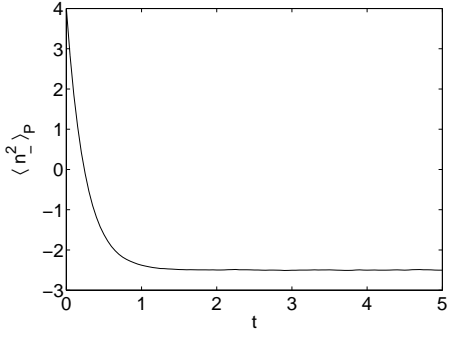}

\caption{Sampled moments of $\langle n_{-}^{2}\rangle_{P}$ , for genetic mutation example in the Poisson representation, parameters
as in text. Sampling error is of order $10^{-3}$ or less. Negative values indicate sub-Poissonian statistics due to mutations
that cause strong inter-species correlations. To this graphical resolution, the exact and numerically simulated results are
indistinguishable.\label{cap:Sampled-moments-of}}
\end{figure}

Fig (\ref{cap:Sampled-mean-populations}) shows the mean values obtained from numerical simulation of these equations for
an ensemble of $10^{6}$ trajectories, showing that the exact analytic result is compatible with the upper and lower one
standard deviation error bounds from the simulations. Results for cross-correlations are shown in Fig (\ref{cap:Sampled-moments-of}),
also agreeing extremely well with the analytic predictions. The numerical and analytic results are indistinguishable at this
graphic resolution. 

The stochastic equations were integrated for a total time of $t=5$, using values of $k=k_{m}=1$. The minimum step-sizes
used were $\Delta t=0.01$ and $\Delta t=0.005$ (to enable a check on the errors due to finite time-steps, which were negligible).
Initial values were set to $n_{+}=5$, $n_{-}=2$, to give results in low population regions with large departures from Poissonian
behavior. All simulation results agree well within the sampling error at $t=5$, as shown in Table (\ref{cap:Table-of-observed}).

\begin{table}
\begin{tabular}{|c|c|c|}
\hline 
Moment&
Analytic&
Poisson\tabularnewline
\hline
\hline 
$\langle n_{+}\rangle_{P}$&
$5.0$&
$5.002(7)$\tabularnewline
\hline 
$\langle n_{-}\rangle_{P}$&
$0.908\times10^{-4}$&
$0.908(0)\times10^{-4}$\tabularnewline
\hline 
$\langle n_{+}^{2}\rangle_{P}$&
$75$&
$75.1(2)$\tabularnewline
\hline 
$\langle n_{-}^{2}\rangle_{P}$&
$-2.5$&
$-2.502(7)$\tabularnewline
\hline 
$\langle n_{+}n_{-}\rangle_{P}$&
$0.4540\times10^{-3}$&
$0.4541(6)\times10^{-3}$\tabularnewline
\hline
\end{tabular}

\caption{Table of observed moments, comparing analytic and simulated results for the genetic mutation equations in the Poisson expansion
at $t=5$. Standard deviations $\sigma_{g}$ for the last significant digit are in brackets. These results for mean populations,
variances and correlations demonstrate that in this case, the Poisson stochastic equations agree with the known analytic
results within the sampling error of the simulations.\label{cap:Table-of-observed}}
\end{table}

In summary, all the simulation results are in excellent agreement with analytic predictions for this model. No boundary term
errors are found using these linear equations, as one might expect, since there is no possibility of movable singularities
with linear drift equations.

\section{Astrophysical molecular hydrogen production}

Stochastic gauges are only needed when the equations are nonlinear, which comes about when multi-component competition or
formation processes are present. To give a typical example of this, consider the astrophysically important problem of hydrogen
recombination to form molecules on interstellar grain surfaces\cite{Biham}. This is thought to be the major source of interstellar
$H_{2}$, and it is known that conventional rate equations are unable to describe this accurately, due to low occupation
numbers at the critical step of dimer formation. The main reactions are:\begin{eqnarray}
H^{(IN)} & \,^{R}\, & H\nonumber \\
2H & \,^{\underrightarrow{k_{1}}}\, & H_{2}\nonumber \\
2H & \,^{\underrightarrow{k_{2}}}\, & H_{2}^{*}\nonumber \\
H & \,^{\underrightarrow{\gamma}}\, & H^{*}\nonumber \\
H_{2} & \,^{\underrightarrow{\gamma_{2}}}\, & H_{2}^{*}\,\,\,.\label{eq:Hmaster}\end{eqnarray}

This describes hydrogen atoms $H$ adsorbed onto a grain surface, and forming hydrogen molecules $H_{2}$ on the grain. The
number of adsorbed atoms can grow via a generation rate ($R$) from an input flux $H^{(IN)}$ , until it reaches an equilibrium.
This occurs due to losses from molecule formation with a total rate of $k=k_{1}+k_{2}$, and from desorption ($\gamma$),
which stabilizes the concentration of $H$ through emission of unbound hydrogen atoms $H^{*}$. The concentration of $H_{2}$
is also stabilized by desorption ($\gamma_{2}$), leading to unbound hydrogen molecules $H_{2}^{*}$. 

There are additional effects due to flux-blocking caused by adsorbed molecules and atoms, as well as dissociation processes
--- which are neglected here for simplicity. Interstellar grains have a distribution of sizes and compositions, which means
that master equations like these need to be solved for a variety of parameter values to give the total molecular production
rate.

Of course, competing processes involving other atomic and molecular species can also occur, leading to an overall situation
of great complexity if all possible molecular species were included. Here I will focus on the elementary case of hydrogen
molecule production. Poisson variables $\alpha_{1},\alpha_{2},\alpha_{3},\alpha_{4}$ can be introduced representing $[H],[H_{2}],[H^{*}],[H_{2}^{*}]$
respectively. In the positive Poisson representation, this leads to the following system of equations:

\begin{eqnarray}
\frac{d\alpha_{1}}{dt} & = & \left[R-\gamma\alpha_{1}-2k\alpha_{1}^{2}\right]+i\alpha_{1}\sqrt{2k}\zeta(t)\,\nonumber \\
\frac{d\alpha_{2}}{dt} & = & k_{1}\alpha_{1}^{2}-\gamma_{2}\alpha_{2}\nonumber \\
\frac{d\alpha_{3}}{dt} & = & \gamma\alpha_{1}\,\nonumber \\
\frac{d\alpha_{4}}{dt} & = & k_{2}\alpha_{1}^{2}+\gamma_{2}\alpha_{2}\,\,.\end{eqnarray}

It should be noted that the first equation can be solved independently from the other ones --- one can also show this at
the level of the Fokker-Planck equation, by simply integrating out all the other variables. The astrophysical molecular production
rate of interest is:

\begin{equation}
R_{H_{2}^{*}}=\langle k_{2}\alpha_{1}^{2}+\gamma_{2}\alpha_{2}\rangle\,\,.\end{equation}

In this model, all hydrogen molecules created are eventually desorbed, since in the steady-state $\langle k_{1}\alpha_{1}^{2}\rangle=\langle\gamma_{2}\alpha_{2}\rangle$.
Hence, the total molecular production rate in the steady-state is obtainable from the solution to the first equation:\begin{eqnarray}
R_{H_{2}^{*}} & = & k\langle\alpha_{1}^{2}\rangle\nonumber \\
 & = & k\langle N_{1}(N_{1}-1)\rangle\,\,.\end{eqnarray}

The important point of physics here is that the hydrogen molecule production rate is proportional to the auto-correlation
function of the hydrogen atom density --- and hence requires a knowledge of the correlations and fluctuations present. This
of course, has a simple physical origin, since hydrogen molecules can only form if at least two atoms are present simultaneously.

\subsection{Analytic solutions}

For notational simplicity, I define $n=\alpha_{1}$, which is the Poisson variable that correspond to the hydrogen atom number.
From the Poisson expansion viewpoint, the only non-trivial term is the hydrogen equation, as this introduces noise. All the
other equations can be solved once the hydrogen number fluctuations are known. 

It is useful to obtain the steady-state hydrogen fluctuations from the complex Poisson representation defined in Eq (\ref{ComplexP}),
as this has an analytic solution for the steady state. The reduced Fokker-Planck equation for the hydrogen atom variables
is simply:\begin{equation}
\frac{\partial}{\partial t}f(n,t)=\left[\frac{\partial}{\partial n}\left(-R+\gamma n+2kn^{2}\right)-k\frac{\partial}{\partial n^{2}}n^{2}\right]f(n,t)\,\,.\end{equation}

This has a steady-state which is exactly soluble, though defined on a complex contour starting and ending at the origin:\begin{equation}
f(n,\infty)=Cn^{(\gamma/k-2)}\exp\left(2n+\frac{R}{kn}\right)\,\,.\label{COMPLEXSOLN}\end{equation}

Here I have kept the derivatives in analytic form, to obtain the simplest potential solution. However, the result is instructive,
since it is clear that this analytic form is inherently complex. This is the essential reason why a gauge variable $\Omega$
is useful in order to get simulations that behave like this simple, compact solution. The gauge variable can attain complex
values during a stochastic calculation, even when the distribution itself is constrained to have positive values. 

In the case of complex valued solutions as in Eq (\ref{COMPLEXSOLN}) it is necessary to choose an appropriate integration
contour to define the manifold over which the analytic derivatives are defined. For simplicity, I introduce relative flux
and relaxation parameters, $\varepsilon=R/(2k)$ and $\rho=\gamma/2k$. Next, using a Sommerfeld contour-integral identity
in the inverse variable $z=1/n$ , one obtains the result for the moments that:\begin{eqnarray}
\left\langle n^{m}\right\rangle  & =C & \int_{-\infty}^{(0+)}z^{(2-m-2\rho)}e^{2\varepsilon(z+1/(\varepsilon z))}dz\nonumber \\
\nonumber \\ & = & \varepsilon^{m/2}I_{2\rho+m-1}(4\sqrt{\varepsilon})/I_{2\rho-1}(4\sqrt{\varepsilon})\,\,.\label{eq:moment}\end{eqnarray}
This exact solution gives the steady-state $H_{2}^{*}$ production rate (neglecting dissociation): $R_{H_{2}^{*}}=k\left\langle n^{2}\right\rangle $
. 

An obvious result, coming from the asymptotic properties of Bessel functions, is that \begin{eqnarray}
\lim_{\varepsilon\rightarrow\infty}\left\langle n^{m}\right\rangle  & = & \varepsilon^{m/2}\nonumber \\
\lim_{\varepsilon\rightarrow0}\left\langle n^{M}\right\rangle  & = & \frac{R^{m}}{(\gamma+k[m-1])\times\ldots\times(\gamma)}\,\,.\label{Asymppt}\end{eqnarray}
Thus for large grains with $\varepsilon\rightarrow\infty$ the high flux limit is just $R_{H_{2}^{*}}=R/2$, which is also
the rate-equation limit. At low fluxes (i.e., small grains) a dramatic and physically understandable feature is obtained:
the $H_{2}^{*}$ production rate can be suppressed below the rate equation result. In this limit of $\varepsilon\rightarrow0$
,\begin{eqnarray}
\left\langle n\right\rangle  & = & \frac{R}{\gamma}\nonumber \\
\left\langle n^{2}\right\rangle  & = & \frac{R^{2}}{\gamma(k+\gamma)}\,\,.\label{Lowflux}\end{eqnarray}

For $k\gg\gamma$, this predicts enormously reduced hydrogen molecule production rates compared to normal rate equations.
The reason for this is simply that when there is only one atom at a time on the grain, no molecules are produced. Similar
results have been found in earlier Monte Carlo calculations as well\cite{Biham}.

\subsection{Poisson equations}

It is simplest to use a scaled time $\tau=2kt$ to calculate the stochastic equations in the Poisson representation for $n=\alpha_{1}$
. With this variable the drift and noise matrices are both scalars; the resulting Ito equations of motion are \emph{unstable
in the absence of gauge terms}: 

\begin{equation}
\frac{dn}{d\tau}=\left[\varepsilon-\rho n-n^{2}\right]+in\eta(\tau)\,\,,\end{equation}
where $\langle\eta(\tau)\eta(\tau')\rangle=\delta(\tau-\tau')$. 

There is a singular trajectory $n\rightarrow-\infty$ which can be accessed via the complex diffusion of $n$ into the negative
half-space of $n<0$ ; this is the instability already encountered in the solutions to the dimerization equation (\ref{dimerizeP}).
For numerical purposes, it is advantageous to use the Stratonovich form, suitable for central difference algorithms:\begin{equation}
\frac{dn}{d\tau}=\varepsilon-n\left[\rho-1/2+n\right]+in\eta(\tau)\,\,.\end{equation}

\begin{figure}
\includegraphics[%
  width=6cm,
  keepaspectratio]{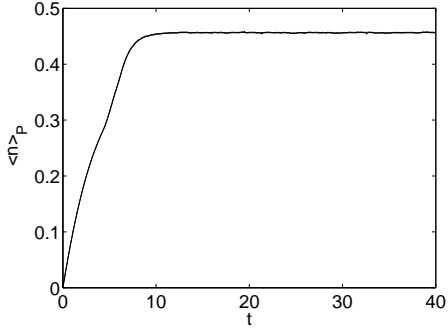}

\caption{Sampled moments of $\langle n\rangle$ for astrophysical hydrogen molecule production in the Poisson representation, parameters
as in text. Adjacent lines give upper and lower ($\pm\sigma_{g})$ error bounds caused by sampling error.\label{Poisson-moments-of-n}}
\end{figure}

\begin{figure}
\includegraphics[%
  width=6cm,
  keepaspectratio]{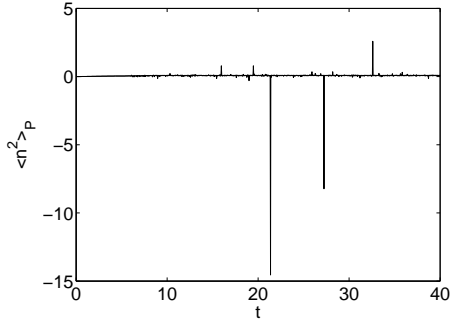}

\caption{Sampled moments of $\langle n^{2}\rangle$ for astrophysical hydrogen molecule production in the Poisson representation,
parameters as in text. Adjacent lines give upper and lower ($\pm\sigma_{g})$ error bounds caused by sampling error.\label{Poisson-moments-of-n^2}}
\end{figure}

Results of the numerical simulations of the Stratonovich equations for hydrogen molecule formation problem in the standard
Poisson representation, showing upper and lower one-standard deviation error-curves, are given in Fig (\ref{Poisson-moments-of-n})
and (\ref{Poisson-moments-of-n^2}). The results clearly show the problems caused by the dynamical instabilities in these
equations, which cause both a large sampling error, especially in $\langle n^{2}\rangle$, as well as systematic errors.
This is especially noticeable in $\langle n\rangle$, which has a relatively low sampling error, and is systematically incorrect.
The steady-state value for these parameters is $\langle n\rangle=0.407..$, which disagrees with the simulations by a margin
much larger than the measured sampling error. The large sampling error in $\langle n^{2}\rangle$ is exactly what is expected
from the inverse power law distribution tails, which mean that the standard deviation in this moment is undefined.

These equations are difficult to integrate numerically, owing to the instabilities, and it is essential to integrate by alternating
between $n$ (for $|n|<1$ , and $z=1/n$ (for $|n|>1$) in order to obtain stable numerical results. Numerical results throughout
this section were obtained using parameters of $\varepsilon=0.1$, $\rho=0.1$ for which the analytic results can be easily
calculated from the Bessel function representation. In this region the rate equations break down, and occupation numbers
are very small, which is a testing region of parameter space for these expansions --- since the fluctuations are far from
Poissonian. The integrations were for a total time of $\tau=t=40$ to allow an approximate numerical steady-state to be reached
from an initial value of $n=0$ (for simplicity, a value of $k=1/2$ was taken). The minimum step-sizes used were $\Delta t=0.005$
and $0.0025$.

\subsection{Stochastic gauges}

Fortunately, it is simple to stabilize these equations by adding non-analytic corrections to the drift. From the basic stochastic
gauge equations (\ref{eq:GSDE}), with a scalar gauge $g$, the resulting Ito equations for astrophysical hydrogen production,
are:

\begin{eqnarray}
\frac{d\Omega}{d\tau} & = & \Omega g\eta(\tau)\,\nonumber \\
\frac{dn}{d\tau} & = & \left[\varepsilon-\rho n-n^{2}\right]+in[\eta(\tau)-g]\,\,.\label{eq:Ito}\end{eqnarray}

For example, consider the effects of three different gauges which all stabilize the equations. The first two correspond to
the amplitude {[}a{]} and phase {[}p{]} gauges treated in the previous section, described by Eq (\ref{amplitudegauge}) and
Eq (\ref{phasegauge}) respectively. The third one is another stabilizing gauge which only acts in the left half-space of
$\Re(n)<0$, where the instabilities are located in this example. This is called the `step' {[}s{]} gauge. 

Defining $n=x+iy,$ the three stabilizing gauges considered are:

\begin{eqnarray}
g_{a} & = & i(n-|n|)\,\,\,\,[a]\nonumber \\
g_{p} & = & i(x-|n|)\,\,\,\,[p]\nonumber \\
g_{s} & = & 2ix\theta(-x)\,\,[s]\,\,.\label{eq:3gauges}\end{eqnarray}

Noting that here $B=in$, these give rise to the following three Ito equations in phase space, each of which is manifestly
stable at large $|n|$ :\begin{eqnarray}
\frac{dn}{d\tau} & = & \varepsilon-n\left[\rho+|n|\right]+in\eta(\tau)\,\,\,[a]\nonumber \\
\frac{dn}{d\tau} & = & \varepsilon-n\left[\rho+|n|+iy\right]+in\eta(\tau)\,\,\,[p]\nonumber \\
\frac{dn}{d\tau} & = & \varepsilon-n\left[\rho+|x|+iy\right]+in\eta(\tau)\,\,\,[s]\,\,.\end{eqnarray}

For numerical integration, it is more efficient to transform to the Stratonovich form, and of course the gauge weight equations
are necessary for weighting purposes. In the amplitude gauge, the Stratonovich equations are:\begin{eqnarray}
\frac{d\Omega}{d\tau} & = & \Omega\left[g_{a}\,\eta(\tau)+(n-g_{a}^{2})/2\right]\nonumber \\
\frac{dn}{d\tau} & = & \varepsilon-n\left[\rho-1/2+|n|\right]+in\eta(\tau)\,\,.\label{eq:Strat}\end{eqnarray}
 In the phase gauge, the Stratonovich equations are:

\begin{eqnarray}
\frac{d\Omega}{d\tau} & = & \Omega\left[g_{p}\,\eta(\tau)+(iy-g_{p}^{2})/2\right]\nonumber \\
\frac{dn}{d\tau} & = & \varepsilon-n\left[\rho-1/2+|n|+iy\right]+in\eta(\tau)\,\,.\label{eq:StratPhase}\end{eqnarray}
In the step gauge, the equations are :\begin{eqnarray}
\frac{d\Omega}{d\tau} & = & \Omega\left[g_{s}\,\eta(\tau)+iy-g_{s}^{2}/2\right]\theta(-x)\nonumber \\
\frac{dn}{d\tau} & = & \varepsilon-n\left[\rho-1/2+|x|+iy\right]+in\eta(\tau)\,\,.\label{eq:Step}\end{eqnarray}

Clearly the first two equations only have inward drift vectors with $d|n|/d\tau<0$ at large enough $|n|$. The last, the
step gauge, is similar, except that it shows linear growth if $\rho<1/2$ and $x=0$. At worst this can only lead to a singularity
in infinite time, and in any event the $y$-axis is not an attractor: so growth along the $y$ axis only leads to a temporary
increase in radius, not a singularity. Hence, all three gauges are completely stable, with no movable singularities.

\subsection{Numerical simulations}

As results in all three gauges were similar, apart from changes to the sampling error, I will only show graphs of the detailed
results in the phase-stabilized gauge, using the same parameter values as previously.

\begin{figure}
\includegraphics[%
  width=6cm,
  keepaspectratio]{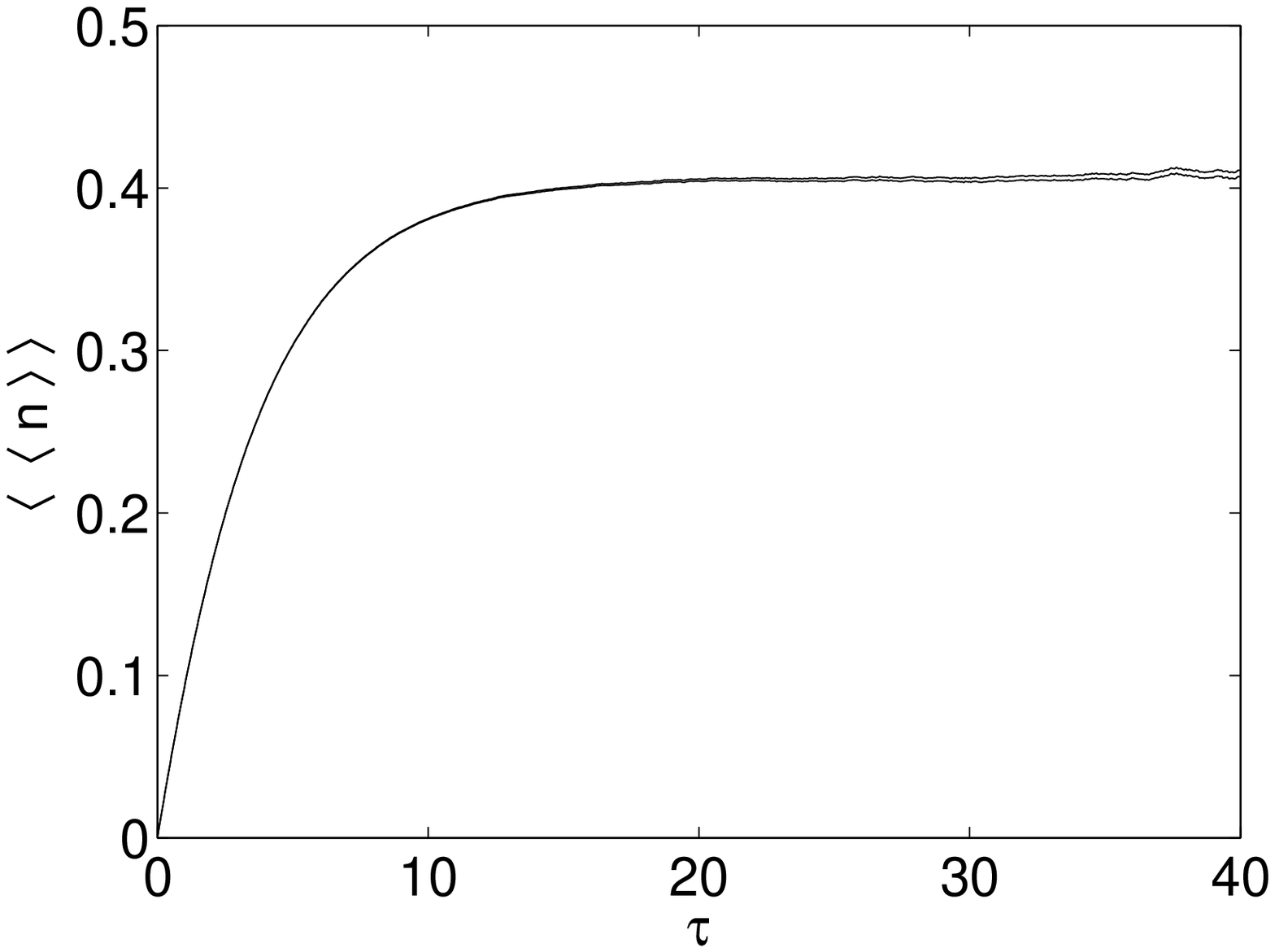}

\includegraphics[%
  width=6cm]{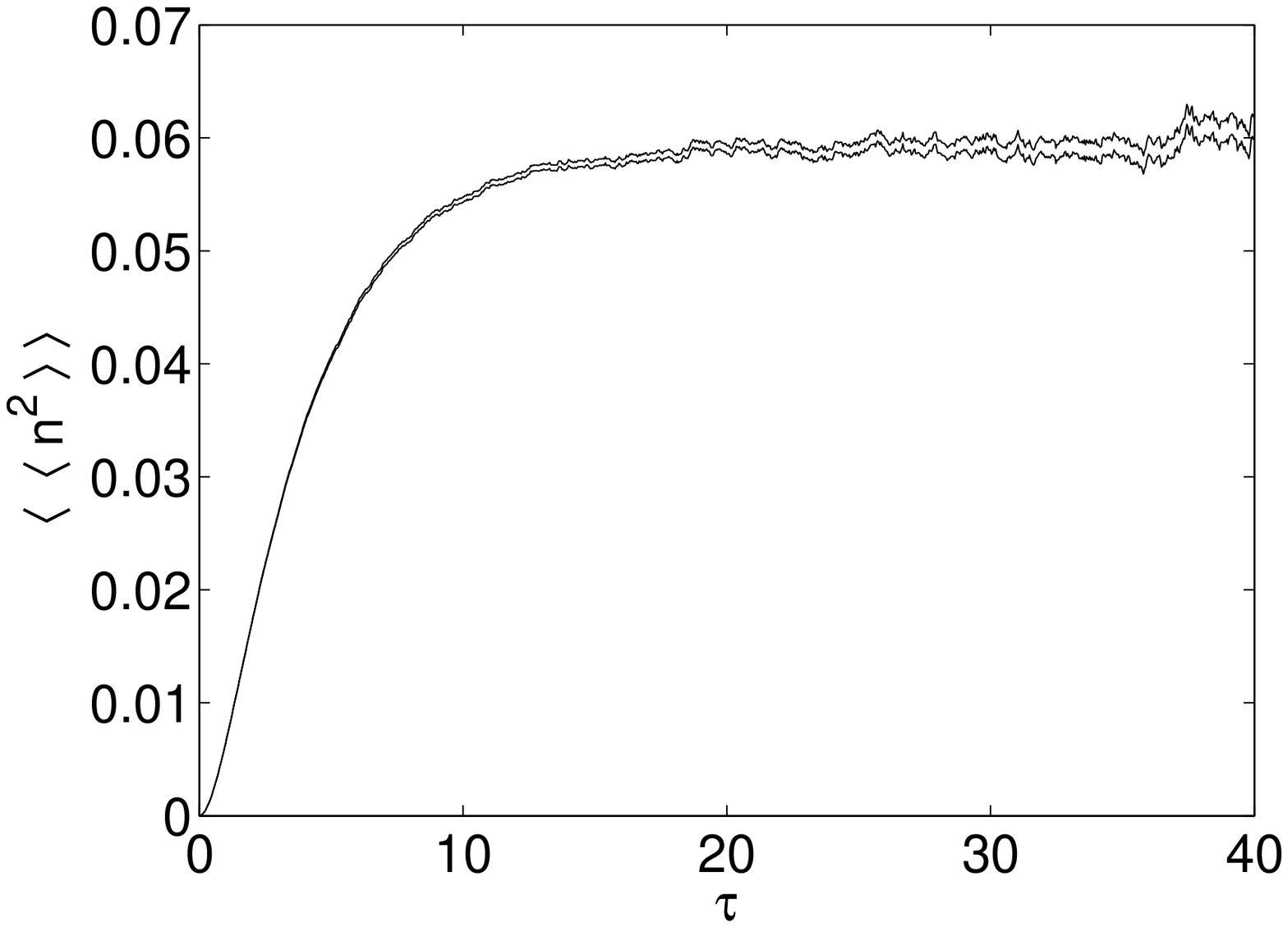}

\caption{Sampled moments of $\langle\langle n\rangle\rangle$ (upper plot) and $\langle\langle n^{2}\rangle\rangle$ (lower plot)
for astrophysical hydrogen molecule production in the `phase' gauge, parameters as in text. Adjacent lines give upper and
lower ($\pm\sigma_{g})$ error bounds caused by sampling error.}
\end{figure}

Results of the numerical simulations in the phase-stabilized gauge, showing upper and lower one-standard deviation error-curves,
are given in Fig (5). It is clearly dramatically improved compared to the Poisson results. 

\begin{figure}
\includegraphics[%
  width=6cm,
  keepaspectratio]{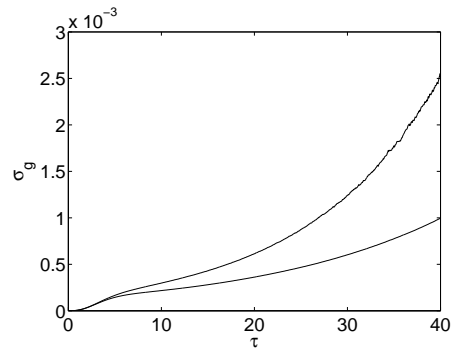}

\caption{Sampling errors: standard deviation $\sigma_{g}$ in the mean of $\langle\langle n^{2}\rangle\rangle$ for astrophysical
hydrogen in the phase gauge (lower curve) , and amplitude gauge (upper curve) .}
\end{figure}
Fig (6) shows that there are reductions of up to four orders of magnitude in the sampling error of molecule production rates,
relative to the Poisson method.

\subsection{Comparison of moments and sampling errors}

Apart from the unmodified Poisson or `zero gauge' results, the gauge simulations are stable. Nevertheless, on closer inspection,
the stable gauges don't behave in an identical way as regards the sampling error with a finite set of trajectories. This
can be seen from the previous figure, which compares two stable gauges.

\begin{table}
\begin{tabular}{|c|c|c|c|c|c|}
\hline 
Moment&
Analytic&
Poisson&
Phase&
Amplitude&
Step\tabularnewline
\hline
\hline 
$\langle\Omega\rangle$&
$1.0$&
$1.0$&
$1.003(4)$&
$0.993(10)$&
$1.005(6)$\tabularnewline
\hline 
$\langle\langle n\rangle\rangle$&
$0.407\ldots$&
$0.456(4)$&
$0.409(2)$&
$0.399(5)$&
$0.406(4)$\tabularnewline
\hline 
$\langle\langle n^{2}\rangle\rangle$&
$0.059\ldots$&
$0.077(5)$&
$0.061(1)$&
$0.058(2)$&
$0.064(3)$\tabularnewline
\hline
\end{tabular}

\caption{Table comparing analytic and simulated results for three different stochastic gauges and the Poisson expansion; the moment
$\langle\langle n^{2}\rangle\rangle$ is critical for molecule production. Sampling error ($\sigma_{g}$) in brackets.\label{cap:Table-comparing}}
\end{table}

For each gauge and for the Poisson expansion, the observed moment and its sampling error $\sigma_{g}$ (standard deviation
in the mean) is given in Table (\ref{cap:Table-comparing}), which tabulates the final near-equilibrium simulation results
at $\tau=40$, and compares them to the equilibrium analytic result for $\tau=\infty$. For the stable gauges, the results
are within $\sigma_{g}$ of the analytic calculations in most cases, and are within $2\sigma_{g}$ in the remaining more
accurate cases --- where the residual discrepancy was partly due to the finite time-step discretization error of around $\pm10^{-3}$.
This indicates that all these (stable) gauges converge to the analytically known correct answer. 

The corresponding (unstable) Poisson method clearly gives incorrect answers due to boundary term and/or sampling errors,
with up to $12\sigma_{g}$ discrepancy in the case of the mean number of hydrogen atoms, $\langle n\rangle$. The graphical
and tabular evidence indicates that the mean atom number is incorrect because the unstable trajectories cause power-law tails
in the distribution, and consequent boundary term errors. In addition, the graph shows that the Poisson time-history has
large fluctuations with sampling errors of up to $1000\%$, showing no signs of equilibration for the molecule production
rate, which is proportional to $\langle n^{2}\rangle$. This is further evidence for power-law tails, which are also found
in a similar quantum-optical master equation.

The amplitude gauge has no systematic errors, but gives the worst sampling error of the stable gauges, as the $n$ variable
is the least constrained in this gauge, tending to diffuse in a circle. For these parameters the phase gauge gives the best
results, as it localizes the $n$ variable near a deterministic stable point. The last gauge is a step gauge --- only giving
non-zero corrections when $x<0$. This has the feature that the gauge term $\Omega$ only changes when the trajectory reaches
$x<0$, and gives sampling errors intermediate between the others.

One might expect that the step gauge would give lower sampling errors in $\langle\Omega\rangle$, since this gauge is zero
in the right half-plane. Instead, the phase-stabilized gauge gives the lowest overall sampling errors for all quantities
with these parameter values, even for the gauge amplitude $\langle\Omega\rangle$ . This is an example of `prevention is
better than cure'. That is, phase-stabilization is also able to prevent amplitudes from growing along the $\pm y$ axis.
The step gauge corrects this growth too late for optimal results, having to use a numerically bigger gauge correction ---
with larger sampling errors.

\section{Conclusion}

The gauge Poisson method is shown to generate a stochastic differential equation that is exactly equivalent to a nonlinear
master equation in certain cases. By comparison, the system-size expansion is only approximate, and the positive Poisson
representation is not exact for problems which have boundary terms due to movable singularities. The gauge technique provides
a way to eliminate boundary-term errors due to singular trajectories. The price paid for this advantage is an extra stochastic
gauge amplitude, which generates a sampling error that grows in time. The focus of numerical simulations in this paper is
on cases where the existence of exact analytic results allows the issue of random and systematic errors to be carefully investigated.
While this is not a complete proof that boundary terms can be eliminated in all cases, it suggests that choosing a stabilizing
gauge is a necessary condition.

This type of model is very general. For example, one can easily include linear diffusion and extend the theory to treat fluctuations
in reaction-diffusion models or even Boltzmann kinetics\cite{Poisson}. The method simply requires that the populations are
defined as occurring in a lattice of bounded cells, either in ordinary space or in a classical phase-space, together with
the appropriate hopping rates. The resulting deterministic equations are just the same as would occur in discretized non-stochastic
equations. However it is necessary to include cell-volume factors in the nonlinear rate constants, so that the noise terms
vary with the $j-$th lattice cell volume $V_{j}$- typically resulting in stochastic noise terms proportional to $1/\sqrt{V_{j}}$.
Applications to these problems will be treated elsewhere.

The technique can be easily extended to include a large number of coupled kinetic processes, as occurs both in genetics,
and in the generation of chemical species of astrophysical importance: for example, $OH,$ $H_{2}O$ , $CO$ and so on. By
contrast, the direct solution of the master equation grows exponentially more complex as the number of interacting species
increases. Similar considerations arise when treating biological species with typically very large numbers of genotypes,
or when treating extended spatial (multi-mode) problems. The method of choosing stable gauges developed here may also be
useful for the corresponding quantum problems.

\begin{acknowledgments}
Numerical calculations were carried out using open software from the XMDS project\cite{xmds}. Thanks to the Australian Research
Council and the Alexander von Humboldt-Stiftung for providing support. Useful discussions on genetic models with A. J. Drummond
and on astrophysical models with O. Biham are gratefully acknowledged.
\end{acknowledgments}

\end{document}